\pdfoutput=1
\documentclass[11pt]{article}
\usepackage[final]{acl}

\usepackage{times}
\usepackage{latexsym}
\usepackage[T1]{fontenc}
\usepackage[utf8]{inputenc}
\usepackage{microtype}

\usepackage{caption}
\usepackage{subcaption}
\usepackage{wrapfig}
\usepackage{comment}
\usepackage{blindtext}
\usepackage{graphicx}
\usepackage{pifont} 
\newcommand{\cmark}{\ding{51}}
\newcommand{\xmark}{\ding{55}}

\usepackage{soul}
\usepackage{tipa}
\usepackage{dblfloatfix}
\usepackage{amsmath}
\usepackage{wrapfig,lipsum,array,booktabs,ragged2e}
\usepackage{multirow}
\usepackage{enumitem}
\usepackage{bbold}
\usepackage{amssymb}
\usepackage{rotating}
\usepackage{bbold}
\usepackage{subcaption}
\usepackage{floatrow}
\usepackage{xcolor,colortbl}

\definecolor{valgood}{HTML}{d0e0e3}

\definecolor{valbest}{HTML}{d9ead3}
\newcommand{\valbest}[1]{\colorbox{valbest}{#1}}
\newcommand{\valworst}[1]{\colorbox{red!10}{#1}}
\newcommand{\supervised}[1]{#1{\color[HTML]{CB4335}$\dagger$}}

\title{Measuring and Addressing Indexical Bias in Information Retrieval}

\newcommand{\treelogo}{\raisebox{5pt}{\includegraphics[scale=0.050]{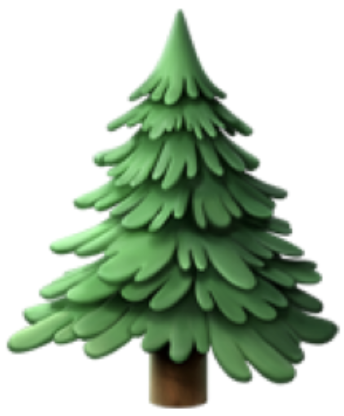}}}
\newcommand{\metalogo}{\raisebox{4pt}{\includegraphics[scale=0.004]{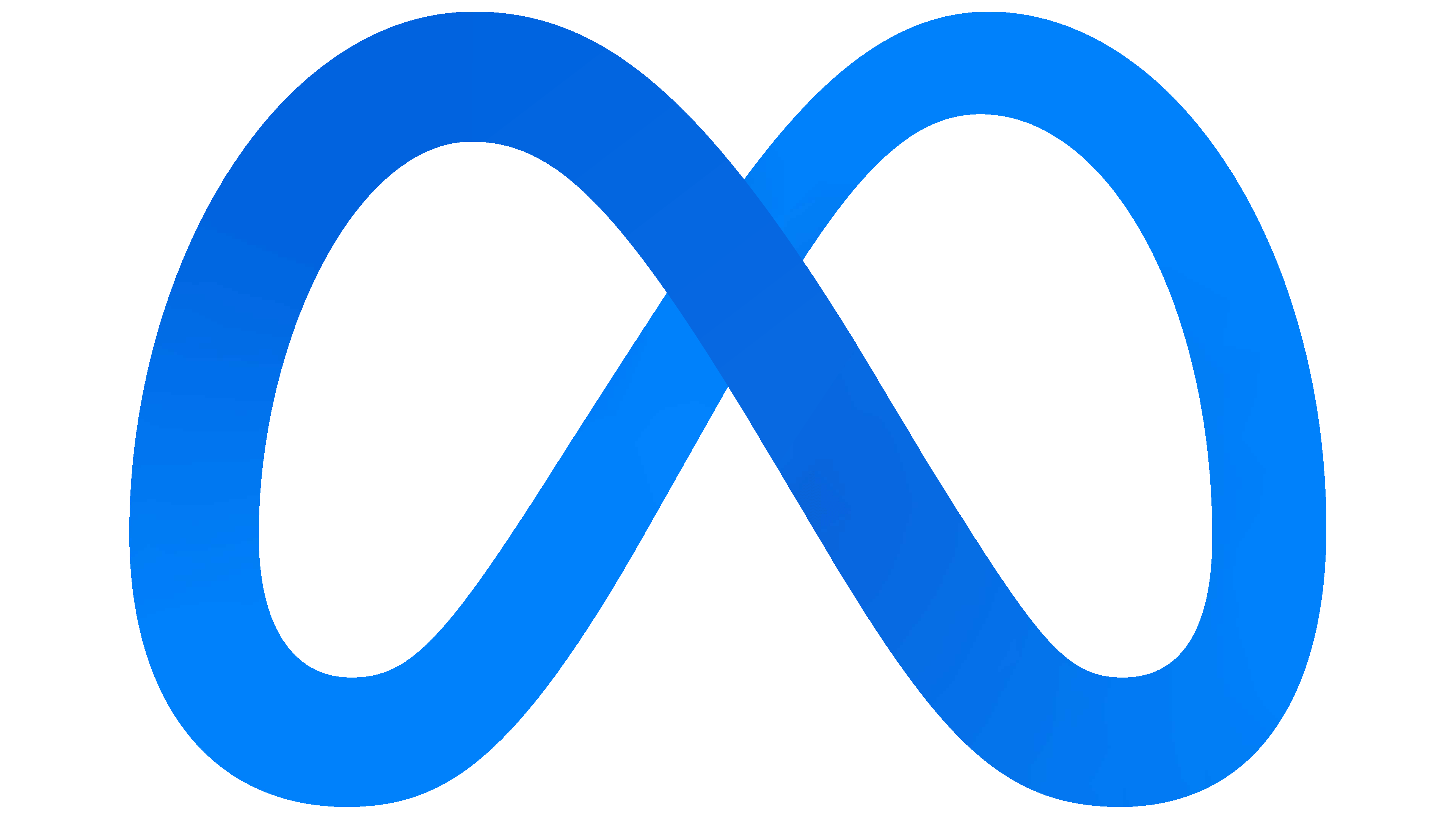}}}
\newcommand{\gtlogo}{\raisebox{3.4pt}{\includegraphics[scale=0.025]{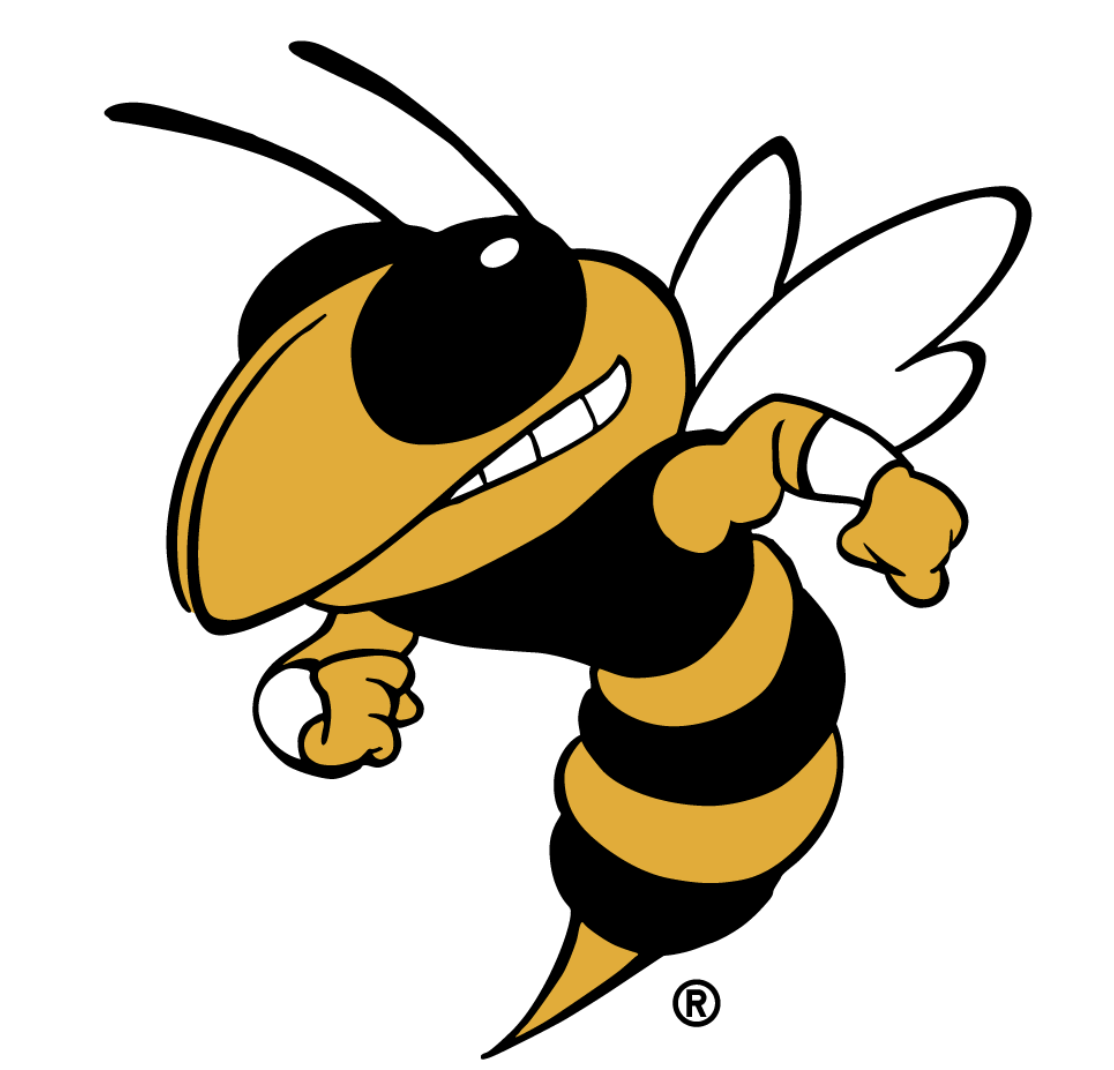}}}
\newcommand{\gt}{\gtlogo}
\newcommand{\stanf}{\treelogo}
\newcommand{\meta}{\metalogo}

\newcommand{\rot}[1]{\begin{turn}{0}#1\enspace\end{turn}}

\newcommand{\numIssues}{1.4k}%
\newcommand{\numDocsSynthetic}{32k} %31534
\newcommand{\numDocsSyntheticGenerated}{138k}%
\newcommand{\numDocsNatural}{4,662}%
\newcommand{\numModelsOpen}{7}%
\newcommand{\numModels}{8}%
\newcommand{\numQueries}{3,996}% 3996 used, 15k available

\newcommand{\numQueriesNatural}{452}
\newcommand{\numBiasDomains}{15}
\newcommand{\numBiasTopics}{1,364}
\newcommand{\numBiasTopicsNatural}{288}
\newcommand{\metric}{\mbox{\textsc{Duo}}}
\newcommand{\pair}{\textsc{Pair}}
\newcommand{\corpus}{\textsc{Wiki-Balance}}
\newcommand{\promptedModel}{\texttt{GPT 3.5 Turbo}}

\author{Caleb Ziems \stanf \hspace{0.3em}
        William Held \gt \hspace{0.3em}
        Jane Dwivedi-Yu \meta \hspace{0.3em}
        \textbf{Diyi Yang} \stanf \\
        \stanf Stanford University, \gt Georgia Institute of Technology, \meta Meta AI \\
        \texttt{\small\{cziems, diyi\}@stanford.edu}, \texttt{\small wheld3@gatech.edu},
        \texttt{\small janeyu@fb.com}
}

\begin{document}
\maketitle
\begin{abstract}
Information Retrieval (IR) systems are designed to deliver \textit{relevant} content, but traditional systems may not optimize rankings for fairness, neutrality, or the balance of ideas. Consequently, IR can often introduce indexical biases, or biases in the positional order of documents. Although indexical bias can demonstrably affect people's opinion, voting patterns, and other behaviors, these issues remain understudied as the field lacks reliable metrics and procedures for automatically measuring indexical bias. Towards this end, we introduce the \pair{} framework, which supports automatic bias audits for ranked documents or entire IR systems. After introducing \metric{}, the first general-purpose automatic bias metric, we run an extensive evaluation of \numModels{} IR systems on a new corpus of \numDocsSynthetic{} synthetic and 4.7k natural documents, with 4k queries spanning \numIssues{} controversial issue topics. A human behavioral study validates our approach, showing that our bias metric can help predict when and how indexical bias will shift a reader's opinion. For data and code, see \url{https://github.com/SALT-NLP/pair}
\end{abstract}

\section{Introduction}

\begin{figure}[t!]
    \centering \includegraphics[width=0.9\linewidth]{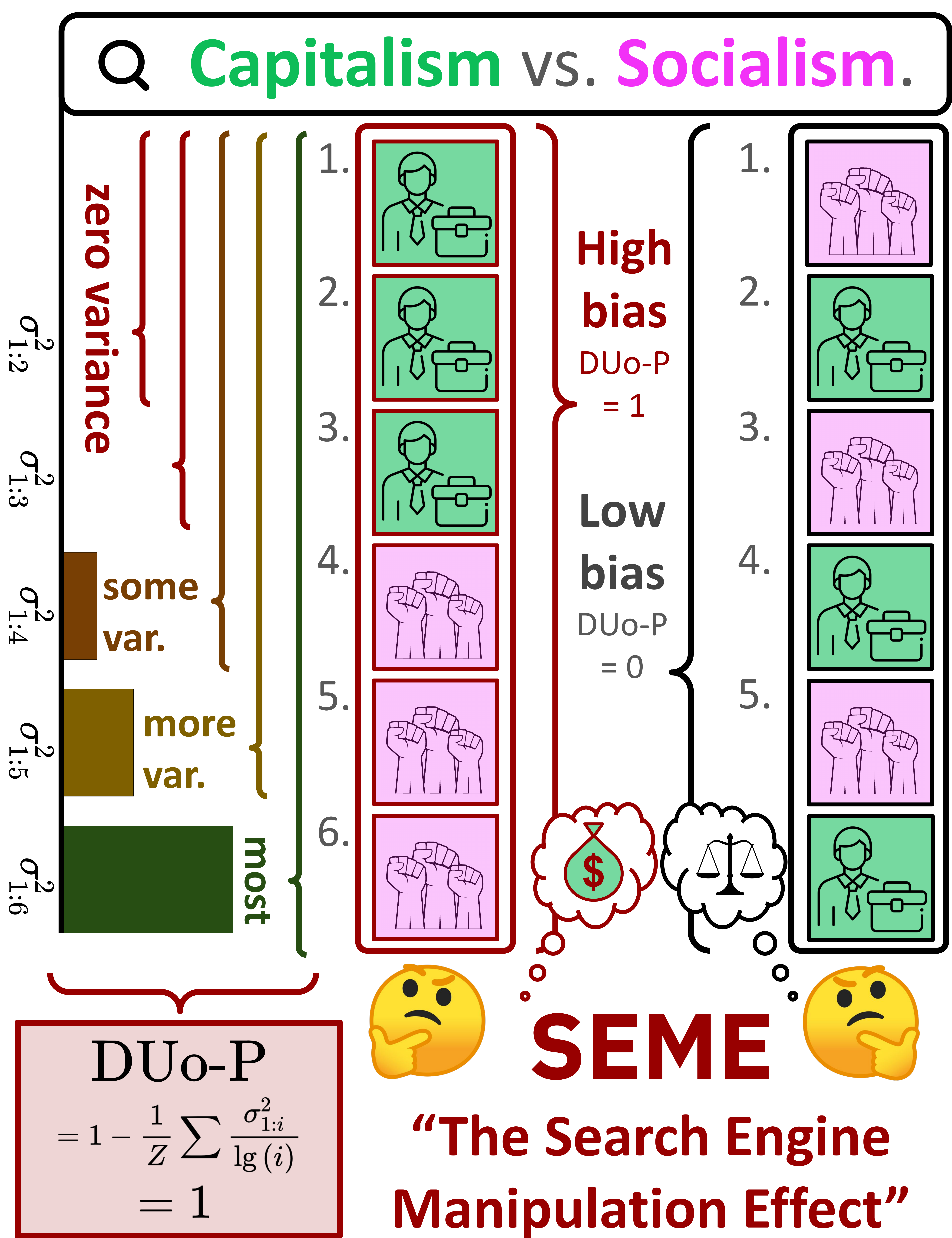}
    \caption{\textbf{The Search Engine Manipulation Effect} as predicted by our \metric{} metric over a set of documents favoring \textit{Capitalism} or \textit{Socialism}. If users read a pro-Capitalism list, they will be more likely to adopt a Capitalist position, and our metric reflects this. The ranking with a most balanced order (\textit{right}) gets the minimal score of \metric{}=0, whereas the documents with the greatest possible indexical bias (\textit{left}) get the greatest score of \metric{}=1. \metric{} uses a discounted sum of variances $\sigma^2_{1:i}$ in polarization across documents' embeddings. On the left, the first 3 \textit{Capitalist} articles have zero variance in polarity, so $\sigma^2_{1:3}=0$. The full list has a variance of $\sigma^2_{1:6}=1$, but since this balance appears far down the ranking, $\sigma^2_{1:6}$ is highly discounted.} 
    \label{fig:crown_jewel}
\end{figure}

Web search, recommendation systems, and personal assistants are all powerful Information Retrieval (IR) tools that can help people make decisions \citep{carroll2014search,mckay2020we}. However, skewed results can lead people to make biased \citep{novin2017making} or misinformed conclusions \citep{bar2009presentation,haas2017ranking}. For example, undecided voters can be swayed to vote for a candidate who is favored in search results \citep{epstein2015search}. This well-known problem is called the search engine manipulation effect \citep[SEME;][]{allam2014impact,azzopardi2021cognitive,draws2021not,epstein2015search,pogacar2017positive}. SEME results not from the content of any particular document, but rather from the \textit{indexical bias} \citep{mowshowitz2002assessing} of their rank order,\footnote{This is also known as \textit{position bias} \citep{biega2018equity}.} since people are more likely to read and trust higher-ranked documents \citep{schwarz2011augmenting}. 

In general, responsible IR should address indexical bias by providing not only relevant content, but also a more fair, balanced, and representative distribution of documents \citep{olteanu2021facts}. To identify and rerank biased results at scale, practitioners need automatic metrics that operate over diverse unlabeled corpora. This motivates \pair{}.

\textbf{P}erspective-\textbf{A}ligned \textbf{I}nformation \textbf{R}etrieval, or \pair{}, is a completely unsupervised method for measuring indexical bias. \pair{} introduces the \textit{\textbf{D}iscounted \textbf{U}niformity \textbf{o}f \textbf{P}erspectives}, or \metric{} bias metric, which measures the variance of perspectives at different ranks within an ordered set of retrieved documents. The \metric{} critically depends on our \corpus{} corpus, which allows us to compute beforehand the most polarized semantic axis of debate for each issue, using principle component analysis over the document embeddings. \metric{} is automatic, unlike prior methods that require human labels. \pair{} is also generalizable, as \corpus{} can be easily expanded beyond the \numIssues{} distinct issues it already supports.

To validate \pair{}, we run a behavioral study which demonstrates how \metric{} can help predict the Search Engine Manipulation Effect (SEME). \metric{} is predictive whenever participants click at least one search result link. In these cases, reranking documents to minimize \metric{} will reduce the SEME. Finally, we leverage \pair{} to evaluate \numModelsOpen{} traditional open-source IR Systems, as well as one commercial search engine. The synthetic \corpus{} corpus serves as a large and representative stress test, revealing specific weaknesses in leading systems. We complement this test with a more natural evaluation over \numDocsNatural{} documents retrieved via Google Search on the same issues. In summary, we contribute:

\begin{enumerate}\setlength\itemsep{0em}
    \item The \metric{} \textbf{Positional Bias Metric}, which is an unsupervised metric that works regardless of the controversial issue of interest.
    \item Two diverse, large-scale \textbf{Bias Evaluation Corpora} with \numDocsSynthetic{} highly polarized synthetic documents and \numDocsNatural{} natural documents.
    \item Extensive \textbf{Bias Audit Evaluations} for \numModels{} IR Systems across \numBiasDomains{} topical domains.
    \item A \textbf{Human Behavioral Study} which validates our approach as predictive of the SEME.
\end{enumerate}

\section{Related Work}

\paragraph{Classifying Stance, Leaning, and Ideology.}
Political leaning, ideology, public opinion, and stance can be tagged at scale with supervised classifiers \cite{johnson2017ideological,luo2020desmog,stefanov2020predicting,baly2020we} or keywords \citep{adamic2005political}. In \pair{}, however, we do not assume access to curated lists or supervised bias classifiers. We also opt not to use zero-shot ideology and stance detection, as LLM performance still varies widely in this domain \citep{ziems2023can}. Instead, we take a fully unsupervised approach and use statistics over generative models.

\paragraph{Diversifying and Debiasing IR.}
Maximum Marginal Relevance \citep[MMR;][]{carbonell1998mrr} is a popular method for diversifying IR results by minimizing documents' mutual similarity. \pair{} embeddings can serve in the MMR similarity metric, but unlike this metric, our \metric{} metric also incorporates rank order. There are also explicit diversification methods such as IA-Select \citep{santos2011intent}, which returns a set with at least one document for each pre-defined category. If such category labels are known in advance, then diversification may be framed as Task-aware Retrieval \citep{asai2022task} and solved with instruction tuning. \citet{zhao2024beyond} find such instruction-tuning methods insufficient for perspective imbalance. Instead they redefine the document-query similarity score to condition on a pre-defined perspective $p$. However, all of these explicit methods may be less generally applicable than \metric{} due to their reliance on document perspective labels, which must be manually-annotated given the limitations above with zero-shot stance detection.

\paragraph{Auditing Bias in IR.}
Numerous prior studies evaluate bias in commercial systems \citep{mowshowitz2002assessing,kay2015unequal,kulshrestha2017quantifying,chen2018investigating,gao2020toward,draws2021assessing} like Google and Bing \citep{gezici2021evaluation}. They often measure the diversity of intents \citep{agrawal2009diversifying,clarke2008novelty,sakai2010simple}, viewpoints \citep{draws2023viewpoint,draws2021assessing}, or fairness towards protected groups \citep{biega2018equity,yang2017measuring,zehlike2022fairness}. Apart from the fairness literature \citep{rekabsaz2020neural},
most prior work runs only case-studies of black-box proprietary search engines, and they rely on bias-keywords \citep{klasnja2022characteristics}, classification, or manual annotation along a particular axis of interest, which is typically binary (e.g., left-right) and centered on American ideologies (e.g., Democrat-Republican). In comparison, \pair{} lets us evaluate open-source IR systems over thousands of distinct issue topics, which can reveal the relationship between system bias and it's underlying algorithm and data. 

\section{Foundational Bias Corpora}
\label{sec:data}

\pair{} relies on two evaluation corpora, \corpus{$_\text{Synthetic}$} and \corpus{$_\text{Natural}$}. The former will critically support the \metric{} computation for automatic, cross-domain evaluation of indexical bias in IR. The latter helps us approximate real-world performance. See Table~\ref{tab:dataset_statistics}, \textit{left} for a comparison, and \textit{right} for statistics. 

\paragraph{Source.} \corpus{} reflects \numBiasTopics{} of the most controversial topics from English Wikipedia, a comprehensive and reliable knowledge source \citep{bruckman2022should}. High-level seed topics come from the titles of Wikipedia articles that were edited in an oscillatory manner (e.g., \textsl{Bullfighting}; \textsl{Beyonc\'e}; \textsl{Climate Change}; \textsl{the Israel-Palestine Conflict}).\footnote{{The subjects of edit wars, NPOV disputes, edit restrictions, or otherwise frequent content revisions, reversions and rollbacks; \url{https://en.wikipedia.org/wiki/Wikipedia:List_of_controversial_issues}}} For details on the topic distribution, see Table~\ref{tab:dataset_statistics} and the discussion in Appendix~\ref{appdx:scope}.

For each topic, we prompt \promptedModel{} to generate 10 specific debate questions on that topic. For example, on \textit{Noam Chomsky}, we generate queries like \textit{Is Noam Chomsky's linguistic theory still relevant?} Figure~\ref{fig:data_pipeline} contains additional examples of \corpus{} queries. For space they are abbreviated in the figure, but all queries have fully grammatical clauses. We sample a subset of these \numDocsSyntheticGenerated{} queries to seed \corpus{}.

\subsection{\corpus{$_\text{Synthetic}$}}
\label{subsec:data_idea_balance}

LLMs generate \corpus{$_\text{Synthetic}$}. For each of \numQueries{} randomly sampled debate questions (e.g., \textsl{Should Karl Marx be considered a revolutionary thinker?}), we create 8 synthetic documents. First, we prompt a \promptedModel{} model with the debate question and: ``\textsl{For each side of the issue, write an article from the perspective of that side.}'' This induces a single polar axis to divide all subsequent documents. 

\begin{figure*}[t!]
    \centering \includegraphics[width=\linewidth]{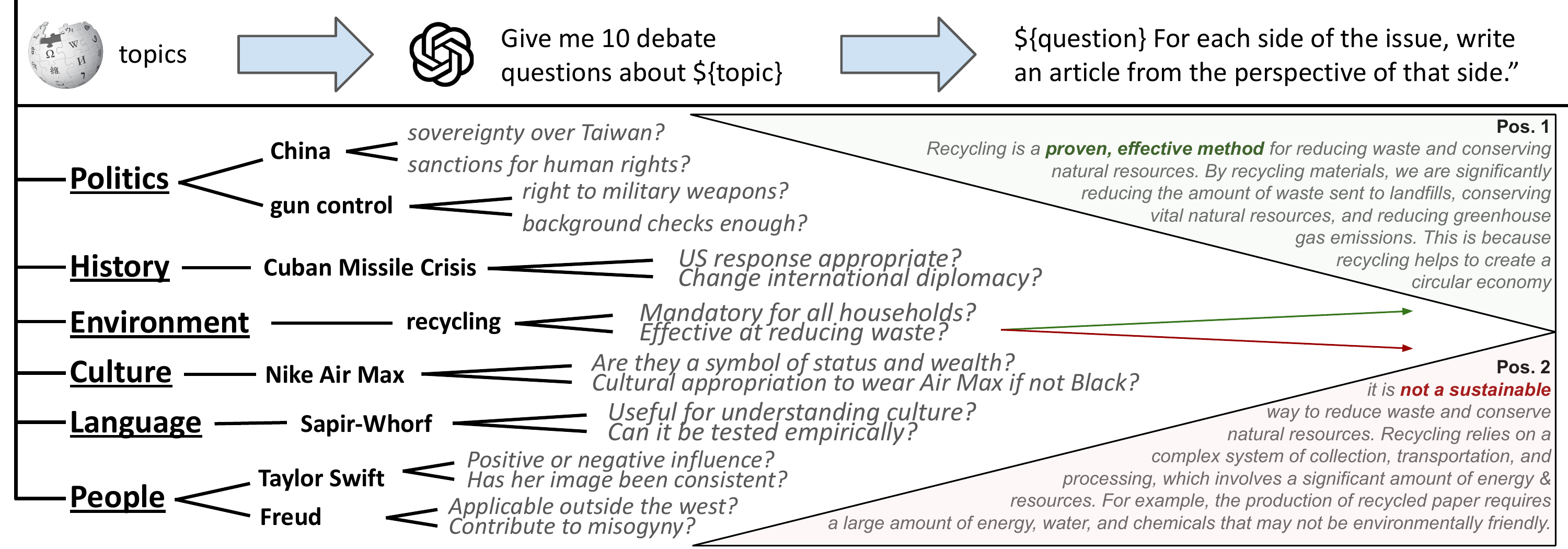}
    \caption{\textbf{\corpus{} Pipeline.} First we pull \numBiasTopics{} controversial topics from English Wikipedia. Data is under CC BY-SA License and is consistent with intended use. Then we generate 10 debate questions about each topic (examples are abbreviated in this figure). For each debate question, we generate 8 polarized documents, with 4 on each side of the initial axis generated by the LLM.}
    \label{fig:data_pipeline}
\end{figure*}

\promptedModel{} gives regular output: each article begins with a header describing the position; e.g., \textsl{\textbf{Perspective 1:} Karl Marx as a Revolutionary Thinker}; \textsl{\textbf{Perspective 2:} Karl Marx as a biased ideologist}). We use regular expressions to extract them, and continue to prompt the model 3 more times: \textit{Given the question above, write an article from the perspective of the side below.}
This produces \numDocsSynthetic{} polarized synthetic documents  (see Figure~\ref{fig:data_pipeline}). In this way, each document is marked as relevant to one out of \numQueries{} queries. Each query in the \corpus{$_\text{Synthetic}$} corpus corresponds to a balanced set of documents, with each half supporting a different perspective.

\subsection{\corpus{$_\text{Natural}$}}
\label{subsec:data_google_search}
This Natural corpus helps us (1) evaluate the Google Search engine, and (2) evaluate open-source IR systems in a real-world setting where documents are less polarized. With \numQueriesNatural{} randomly-sampled queries from \corpus{$_\text{Synthetic}$}, we scrape publicly-available natural web documents from the top 10 results of Google Search in October 2023. To reduce noise, we keep only HTML documents' body text.

\subsection{\corpus{} Quality Estimation} 

\paragraph{Human Evaluation.} To measure the quality of \corpus{$_\text{Synthetic}$}, we recruit domain experts\footnote{For \textsl{Entertainment, History, Religion, and Sports,} we recruit a Graduate Student with a B.S. in Journalism. For \textsl{Politics, Sexuality, Law, and Media,} we find a Public Policy Graduate Student with a B.A. in Political Science. For \textsl{Psychiatry, Technology, and People} we enlist a Nurse in Clinical Behavioral Health with a B.A. in Psychology. And for \textsl{Science and Environment}, we recruit a former CDC health communication specialist with a B.S. in Public Health and an M.S. in Health Education. For \textit{Languages and Philosophy}, we hire a former writing expert at Grammarly with an M.F.A.} from Upwork to blindly evaluate 10-20 random query-document pairs. First, evaluators consider the \textit{subjectiveness} and \textit{topical relevance} of the query. A subjective question does not have a single correct answer, and a relevant query relates to the topic in an interesting and well-specified way. Table~\ref{tab:dataset_statistics} shows that queries are sufficiently subjective and highly relevant to their respective topics. Annotators also score our synthetic documents for faithfulness \citep{durmus2020feqa}, coherence \citep{dang2005overview}, relevance, and fluency, showing that they are high in each of these respects.

\paragraph{Safety.}
Although \corpus{} is centered around controversial issues, we want to reduce risk by measuring and addressing any toxic, harmful, or otherwise unsafe content contained in the documents. Using the OpenAI Moderation API, we determine that no document contains hate, harassment, self-harm, or unwarranted sexual content with a model confidence score larger than 0.09. We manually verified that the most triggering documents were benign cases of almost journalistic reporting on their respective topics (e.g., a debate on the key ingredients of sexual gratification).

\begin{table*}[]
    \centering
    \small
    \resizebox{\textwidth}{!}{%
    \begin{tabular}{rll}
    \toprule
    & \multicolumn{2}{c}{\corpus{}}\\
    & Synthetic & Natural\\ \midrule
    Domains & \numBiasDomains{} & \numBiasDomains{} \\
    Topics & \numBiasTopics{} & \numBiasTopicsNatural{}\\
    Queries & \numQueries{} & \numQueriesNatural{}\\
    Documents & 31,534 & \numDocsNatural{} \\ \midrule
    Google Search & \xmark{} & \cmark{} \\
    Gold Labels & \cmark{} & \xmark{} \\ \midrule
    Applies: rND & \cmark{} & \xmark{} \\
    Applies: rKL & \cmark{} & \xmark{} \\
    Applies: \metric{} & \cmark{} & \cmark{} \\
    \bottomrule
    \end{tabular}
    \begin{tabular}{lrrrlccccccc}
    \toprule
    &&&&& \multicolumn{2}{c}{Query Quality} && \multicolumn{4}{c}{Synthetic Document Quality}\\ \cmidrule{6-7} \cmidrule{9-12} 
    Domain & Topics & Queries & Docs && Relev. & Subj. && Faith. & Coh. & Relev. & Flu. \\
    \midrule
    Entertain. & 26 & 66 & 528 && 4.3 & 3.8 && 5.0 & 5.0 & 5.0 & 4.8\\
History & 122 & 382 & 3,004 && 4.9 & 3.5 && 5.0 & 5.0 & 4.5 & 4.8\\
Law & 15 & 41 & 324 && 3.3 & 4.9 && 4.8 & 5.0 & 4.8 & 5.0\\
Culture & 110 & 323 & 2,550 && 4.5 & 5.0 && 4.8 & 4.8 & 4.7 & 5.0\\
Politics & 243 & 703 & 5,576 && 4.4 & 4.9 && 4.7 & 4.5 & 4.8 & 5.0\\
Religion & 112 & 334 & 2,638 && 5.0 & 3.0 && 4.7 & 4.7 & 4.3 & 4.6\\
Sexuality & 86 & 249 & 1,990 && 4.3 & 5.0 && 4.8 & 5.0 & 5.0 & 5.0\\
Sports & 20 & 57 & 444 && 4.9 & 3.0 && 5.0 & 5.0 & 4.6 & 4.7\\ \midrule
% Environ. & 59 & 164 & 1,286 && - & - && - & - & - & -\\
% Languages & 44 & 109 & 862 && - & - && - & - & - & -\\
% People & 427 & 1,226 & 9,674 && - & - && - & - & - & -\\
% Philosophy & 6 & 23 & 178 && - & - && - & - & - & -\\
% Psychiatry & 4 & 9 & 72 && - & - && - & - & - & -\\
% Science & 139 & 406 & 3,200 && - & - && - & - & - & -\\
% Technology & 33 & 78 & 614 && - & - && - & - & - & -\\ \midrule
    \textbf{Mean} & 91 & 266 & 2,102 && 4.5 & 4.1 && 4.9 & 4.9 & 4.7 & 4.9 \\
    \bottomrule
    \end{tabular}
    }%
    \caption{(\textit{Left}) \textbf{\corpus{} statistics} for both Synthetic and Natural corpora, which both use the same topics and queries, but the latter is much smaller and lacks gold labels, so previous metrics rND and rKL do not apply. (\textit{Right}) \textbf{Quality audit of a random sample of \corpus{} according to human raters.} Humans perceive most queries to be Relevant (Relev.) and sufficiently subjective (Subj.) for use in this task. Synthetic documents are highly Faithful (Faith.), Coherent (Coh.), Relevant to the Query (Relev.) and Fluent (Flu.), which gives us confidence in the validity of the \textsc{Wiki-Bias} resource.}
    \label{tab:dataset_statistics}
\end{table*}

\section{Indexical Bias Metrics}
\label{sec:metrics}

\subsection{Prior Metrics} 
Indexical bias arises when documents of class $A$ receive greater visibility in search results than do documents of class $B$. Due to primacy effects \citep{ho2008estimating}, the document's rank index can serve as a proxy for visibility, as higher-ranked documents will be more visible \citep{joachims2007evaluating,pan2007google,baeza2018bias}, and thus more frequently clicked \citep{insights2013value}. Discounted Cumulative Gain (DCG) assumes a document's visibility, or the attention it receives, is inversely proportional to the logarithm of its index. Although log-based decay may not exactly reflect user attention \citep{ghosh2021fair,sapiezynski2019quantifying}, this has become the standard in IR. DCG is thus defined according to Equation~\ref{eq:dcg}.

{\small
  \setlength{\abovedisplayskip}{6pt}
  \setlength{\belowdisplayskip}{\abovedisplayskip}
  \setlength{\abovedisplayshortskip}{0pt}
  \setlength{\belowdisplayshortskip}{3pt}
\begin{align}
\label{eq:dcg}
\text{DCG}(r, u) = \sum_{i=1}^{|r|}\frac{u(i, r)}{\log_2{i}}    
\end{align}
}
where $u(i,r)$ is the utility of ranking $r$ up to and including document $i$. Researchers can normalize the Discounted Cumulative Gain by setting

{\small
  \setlength{\abovedisplayskip}{6pt}
  \setlength{\belowdisplayskip}{\abovedisplayskip}
  \setlength{\abovedisplayshortskip}{0pt}
  \setlength{\belowdisplayshortskip}{3pt}
\begin{align}
\label{eq:normalization}
 \text{nDCG}(r,u) = \frac{\text{DCG}(r,u) - \min_{r'} \{\text{DCG}(r',u)\}}{\max_{r'} \{\text{DCG}(r',u)\} - \min_{r'} \{\text{DCG}(r',u)\}}   
\end{align}
}%
That is, we set the minimum to zero and divide by the metric's highest attainable value for the given number of items $|r|$ and metric parameters. This means all measures will reach their best, most fair value at 0, and their worst value at 1.

Following the group fairness literature \citep{pedreschi2009measuring,pedreshi2008discrimination}, we can define $u(i,r)$ as statistical parity in the visibility of a protected group \citep{pitoura2022fairness}. In the standard formulation of \citet{yang2017measuring}, they set $\text{rND} = \text{nDCG}(r, u_{ND})$ with
\begin{align}
    \label{eq:nd}
    u_{ND}(i,r) &= {P_{g}@i -P_{g}@\left|r\right|}
\end{align}
which is the difference between the proportion of protected group $g$ members in the top $i$ against the proportion of protected group $g$ members in the population (i.e., the full ranking). The corresponding rND metric is convex and continuous, but not differentiable at zero. To further smooth this metric, \citet{yang2017measuring} also consider the \textit{KL divergence} between protected group membership in the top $i$ vs. the full ranking, giving $\text{rKL} = \text{nDCG}(r, u_{KL})$ with

{\small
  \setlength{\abovedisplayskip}{6pt}
  \setlength{\belowdisplayskip}{\abovedisplayskip}
  \setlength{\abovedisplayshortskip}{0pt}
  \setlength{\belowdisplayshortskip}{3pt}
\begin{align}
    u_{KL}(i,r) &= -P_{g}@i\log_{}\left(\frac{P_g@\left|r\right|}{P_g@i}\right) \nonumber \\ %
    &-\left(1-P_g@i\log_{}\left(\frac{1-P_g@\left|r\right|}{1-P_g@i}\right)\right)
\end{align}
}% 
To apply fairness metrics rND and rKL to the most general case of indexical bias, we can abstract group membership $g$ to indicate whether a document is polarized in a particular direction. We can safely ignore any prior metric for which this generalization would not apply, such as the keyword-bias metric of \citet{rekabsaz2021societal} and the Normalized Discounted Ratio of \citet{yang2017measuring}, which assumes $g$ is a minority group ($P_g@|r|<0.5$). We will also focus entirely on \textit{ranking} bias metrics, and ignore \citet{kulshrestha2019search} and others who measure the bias of individual documents.

\subsection{The \metric{} Bias Metric}
Clearly, rND and rKL can be used only in cases where document polarization labels are known, such as when: (1) we have manually annotated the corpus according to the target axis, or (2) when we generate a controllable synthetic corpus like \corpus{$_\text{Synthetic}$}. Option (1) is not scalable, especially with thousands of distinct axes of controversy in \S\ref{sec:data}. Option (2) does not apply to evaluations in real-world settings. This motivates us to build an unsupervised metric for indexical bias that can operate automatically, even in real-world settings, using scalable knowledge from LLMs. 

Our proposed bias metric is the ``\textit{\textbf{D}iscounted \textbf{U}niformity of \textbf{O}pinions}'' (\metric{}). This unsupervised metric critically depends on our synthetic data to determine the axis of polarization for each query.\footnote{It is important to note that even for our \metric{} computations on the Natural corpus, we rely on \corpus{$_\text{Synthetic}$} here to determine the axis of bias.} Given a topic $t$ represented by a query $q_t$, we pull from \corpus{$_\text{Synthetic}$} a set of $|r|$ synthetic documents that argue each opposing perspective for $q_t$ following \S\ref{subsec:data_idea_balance}. We use a transformer-based document encoder to embed each document into a dense vector, and by running PCA, we project document embeddings into scalar polarization scores $p_j \in \mathbb{R}$, with an average score of $\bar{p} = \frac{1}{|r|} \sum_{i=1}^{|r|} p_j$. This allows us to define
\begin{align}
    \label{eq:polarization_utility}
    u_{V}(i, r) = \frac{1}{i} \sum_{j=1}^{i} (p_j - \bar{p})^2
\end{align}
so that the utility of a ranked subset is the variance of its polarization scores. Now this utility is the complement of a bias metric: more variance in the polarization scores indicates a better balance of ideas and \textit{less} indexical bias. Thus for convenience in this paper, we will refer to our normalized \metric{}$(r)$ metric by the following equation:
\begin{align}
    \label{eq:ndcv}
    \text{\metric{}}(r) = 1 - \text{nDCG}(r, u_V)
\end{align}

\section{Validating \metric{}}
\label{sec:validation}
Experimental evidence demonstrates the validity of the \metric{} metric. Our first experiment shows that the unsupervised polarization score that grounds \metric{} can accurately partition a corpus of polarized documents into their respective viewpoints (\S\ref{subsec:validation_synthetic}). Our second experiment shows that \metric{} can help predict the Search Engine Manipulation Effect in a real behavioral manipulation (\S\ref{subsec:validation_behavioral}). Together, these results demonstrate both the fundamental and psychological validity of \metric{}. 

\subsection{Validation with Synthetic Data}
\label{subsec:validation_synthetic}

Since \metric{} depends on reliable polarization scores to compute viewpoint variance, the first validation step is to evaluate the accuracy of these scores. Accurate polarization scores should partition a set of related documents into contrasting subsets for each viewpoint, where documents with positive scores endorse Perspective 1, and documents with negative scores endorse Perspective 2 for a given query (or vice versa). With viewpoint labels from \S\ref{subsec:data_idea_balance} as ground truth, we compute the maximum accuracy for each query, and average over all queries in \corpus{$_\text{Synthetic}$}. 

Accuracy depends on our document embedding model, so we evaluate all 124 of the models associated with Sentence BERT \citep{reimers2019sentence}, a transformer-based document similarity metric widely used in the IR community. We also evaluate the recent \texttt{voyage-02} embeddings from Voyage AI.\footnote{https://www.voyageai.com/}
Table~\ref{tab:polarization_accuracy_results} provides the worst accuracy (73\%), best accuracy (95\%), and the median (87\%) and mean (86\%) accuracy across all 124 evaluations. These validate our approach since, in the best case, with every third query there is only one incorrect document label among the 8 documents retrieved. For more detailed results, see Tables~\ref{tab:polarization_accuracy_results_above_90} and \ref{tab:polarization_accuracy_results_below_90} in Appendix~\ref{appdx:expanded_polarization_accuracies}.

\begin{table}[t!]
\small
\resizebox{\columnwidth}{!}{%
\def\arraystretch{1.15}
\begin{tabular}{ll}\toprule
\textbf{Embedding Model} & \textbf{Accuracy} \\ \midrule
\textit{Worst:} \texttt{clip-ViT-B-32-multilingual-v1} & 73.19\%\\ 
\textit{Mean Accuracy} & 85.54\% \\
\textit{Median Accuracy} & 85.70\% \\ 
\rowcolor{valbest} \textbf{\textit{Best:} \texttt{sentence-t5-xl}} & \textbf{95.27\%} \\
\bottomrule 
\end{tabular}
}
\caption{\textbf{Polarization score accuracies} for the worst, best, median, and mean performance among all 124 evaluated models. High mean and best performances validate our approach.}
\label{tab:polarization_accuracy_results}
\end{table}

\subsection{Validation with A Behavioral Study}
\label{subsec:validation_behavioral}
\paragraph{Experimental Design.} A behavioral study can demonstrate the psychological validity of the \metric{} metric as predictive of the Search Engine Manipulation Effect (SEME). For each query $q \in \text{\corpus{}}$, a randomly sampled participant will have some opinion $o_\text{prior} \in \mathbb{Z}$. SEME predicts that, after this participant considers a biased list of search results relevant to the given query, their final opinion $o_\text{posterior} \in \mathbb{Z}$ will be shifted with a magnitude proportional to the magnitude of the indexical bias. Our null hypothesis is that \metric{} is unrelated to the effect, so the coefficient $\beta_2$ is zero in the following regression:
\begin{align*}
    o_{\text{posterior}} = \beta_0 + \beta_1 o_{\text{prior}} + \beta_2 \mu^{+}_{\text{\metric{}}} + \epsilon
\end{align*}

To validate \metric{}, we would reject the null hypothesis. Here, $\mu^{+}_{\text{\metric{}}}$ is a signed copy of \metric{} (see Appendix~\ref{appdx:signed_ndcv} for a derivation), which by default measures only the magnitude of the bias and not its direction.
We need a sign to indicate the \textit{direction} of the bias because we expect that shift will move towards the favored perspective.

\begin{figure}[t!]
    \centering \includegraphics[width=\linewidth]{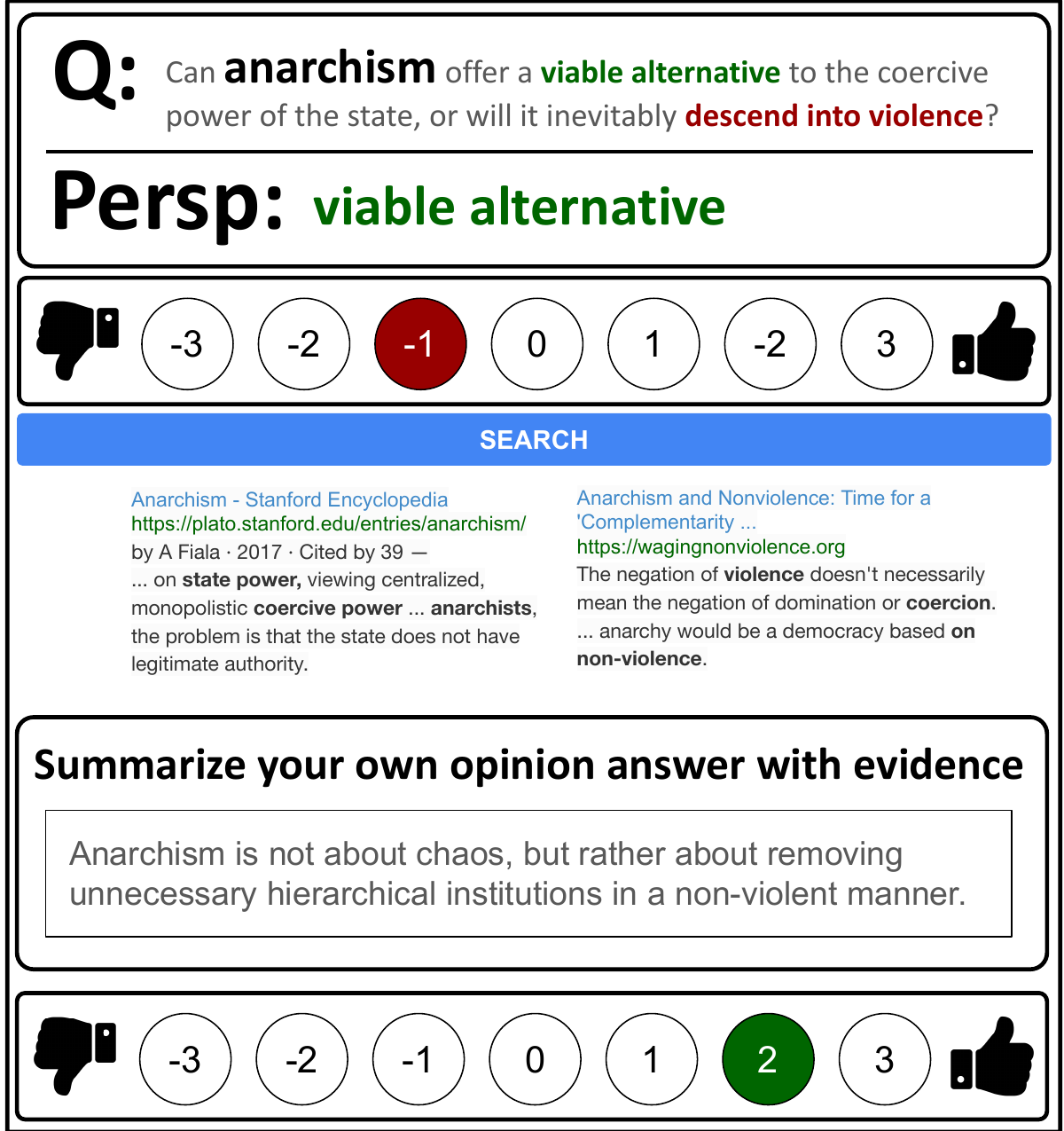}
    \caption{\textbf{Human Behavioral Study Interface} to help determine whether biased search results lead to the SEME. Participants read a query (Q) with a given Perspective (Persp) and tell us whether they agree (3) or disagree (-3) with Perspective. After reading a manipulated list of up to 10 search results, they summarize their informed opinion and provide us their updated agreement on a scale from -3 to 3. We expect more biased results to more radically shift their opinions.}
    \label{fig:behavioral_study}
\end{figure}

\paragraph{Experiment.} 
American adult participants are recruited from Prolific. Each participant interacts with an interface like that shown in Figure~\ref{fig:behavioral_study}. At the top is a query and a given perspective, like \textsl{anarchism is a viable alternative to the coercive power of the state.} The participant then provides their initial opinion on a 7-point Likert scale, $o_\text{prior} \in \{-3, -2, -1, 0, 1, 2, 3\}$, where $-3$ indicates strong disagreement with the listed perspective, and $3$ indicates strong agreement. After entering this prior, the participant uses our manipulated search engine, which retrieves a list of up to 10 search results. Participants are randomly assigned to the experimental manipulation: either results are ordered (1) with the \textit{maximum} bias or (2) \textit{minimum} bias, according to the \metric{} metric. We log any article links the participant clicks, assuming participants are motivated to read these results, since it can assist them in the penultimate task question: summarizing their opinion and quoting evidence. The participant concludes the task after providing $o_\text{posterior}$ on a similar Likert scale. 

\paragraph{Results.} We run our experiment for $N=200$ evaluations on each of our two corpora, as shown in Table~\ref{tab:seme_regression_results}. By logging click behavior, we discover that participants are \textit{not} as often motivated to read search results in the Synthetic corpus as they are in the Natural corpus. With Synthetic, 10\% of users clicked at least one article, while in the Natural, half of users clicked at least one article. Differential click-through behavior effects our findings. When we limit our regression to only those trials where a user clicked at least one link (Behavior: Clicked), we can reject the null hypothesis with statistical significance. So \textbf{only in cases of article click-through, \metric{} significantly helps predict the Search Engine Manipulation Effect} with an $R^2$ effect size greater than 0.48 in both the Natural and Combined corpora, $p<0.05$. These findings are robust, as they replicate with the best and average embedding models (see Table~\ref{tab:seme_regression_results_robustness} in Appendix~\ref{appdx:robustness_checks}). From this, we conclude that \textbf{\metric{} is psychologically valid as it helps predict SEME}. 

\begin{table}[t!]
\small
\resizebox{\columnwidth}{!}{%
\def\arraystretch{1.15}
\begin{tabular}{lrccc|c}\toprule
\textbf{Corpus} & \textbf{Behavior} 
& $N$ & $\beta_2$ & $P(\beta_2=0)$ & \textbf{$R^2$} \\ \midrule
Synthetic & All & 200 & 0.059 & 0.673 & 0.364\\
Synthetic & Clicked & 19 & 0.255 & 0.566 & 0.689\\
Natural & All & 225 & 0.140 & 0.253 & 0.474\\
\rowcolor{valbest} Natural & Clicked & 99 & 0.392 & 0.036 & 0.489\\
\rowcolor{valbest} Combined & Clicked & 118 & 0.365 & 0.032 & 0.519\\
\bottomrule 
\end{tabular}
}
\caption{{\textbf{Regression Results on the Significance of \metric{} in a SEME Behavioral Study} over both the Natural and Synthetic coprora. In natural experiments where participants clicked at least one article link (\valbest{Behavior=\textbf{Clicked}}), we observe significant ($p<0.05$) positive $\beta_2$ coefficients, leading us to conclude that \metric{} helps predict the SEME in cases of article click-through, and thus validating our method.}}
\label{tab:seme_regression_results}
\end{table}

\section{IR Bias Audits}
\label{sec:bias_audits}

\begin{table*}[ht!]
\small
\resizebox{\textwidth}{!}{%
\begin{tabular}{llccccccc|cccc}\toprule
& & \multicolumn{2}{c}{\textbf{Relevance:} \textit{Synthetic}} && \multicolumn{3}{c}{\textbf{Bias:} \textit{Synthetic}} && \multicolumn{2}{c}{\textbf{Relevance:} \textit{Natural}} && \textbf{Bias:} \textit{Natural} \\
\cmidrule{3-4} \cmidrule{6-8} \cmidrule{10-11} \cmidrule{13-13}
Class & Model & nDCG@1 & @10 && rND & rKL & \metric{} && nDCG@1 & @10 && \metric{} \\ \midrule
Lexical & BM-25 & 0.99 & \valbest{0.97} && 0.66 & 0.61 & 0.60 && \valbest{0.87} & \valbest{0.75} && 0.66 \\ \hline
\multirow{2}{*}{Sparse} & SPARTA & 0.99 & 0.95 && \valbest{0.65} & \valbest{0.59} & 0.58 && 0.79 & \valworst{0.65} && 0.65 \\
& SPLADE & \valbest{1.00} & \valbest{0.97} && 0.66 & 0.60 & 0.60 && 0.83 & \valbest{0.75} && \valworst{0.67} \\ \hline
\multirow{3}{*}{Dense} & ANCE & 0.99 & 0.96 && 0.67 & \valworst{0.62} & \valworst{0.62} && 0.81 & 0.71 && \valworst{0.67} \\
& SBERT & 0.99 & 0.96 && \valworst{0.68} & \valworst{0.62} & 0.61 && 0.86 & \valbest{0.75} && 0.65 \\
& Use-QA & \valworst{0.95} & \valworst{0.88} && 0.66 & 0.60 & \valbest{0.57} && \valworst{0.78} & 0.68 && \valbest{0.64} \\ \hline
Late & ColBERT & \valbest{1.00} & \valbest{0.97} && 0.66 & 0.60 & 0.60 && 0.83 & 0.72 && 0.65 \\ \midrule \midrule
Industry & Google &\textit{N/A}\supervised{}& \textit{N/A}\supervised{} && \textit{N/A}\supervised{}& \textit{N/A}\supervised{}& \textit{N/A}\supervised{}&& \textit{N/A}\supervised{}& \textit{N/A}\supervised{}&& \valbest{0.63} \\
\bottomrule 
\end{tabular}
}
\caption{\textbf{Aggregate relevance and bias results} over \corpus{$_\text{Synthetic}$} (\textit{left}) and \corpus{$_\text{Natural}$} (\textit{right})
demonstrate how the most relevant models are not always the least biased. Use-QA and SPARTA have the lowest bias scores, but they are also the least relevant. SPLADE is the most relevant, but also introduces the most indexical bias in the Natural setting. Best results are \valbest{green}, and worst results are \valworst{red}. \supervised{}\textit{N/A} indicates that the metric is not applicable because it requires external human labels.}
\label{tab:aggregate_results}
\end{table*}

\begin{table*}[ht!]
\small
\centering
\resizebox{\textwidth}{!}{%
\def\arraystretch{1.15}
\begin{tabular}{ll|cc|cc|cc|cc|cc|cc}\toprule
Class & Model & \multicolumn{2}{c}{\rot{Entertainment}} & \multicolumn{2}{c}{\rot{Environment}} & \multicolumn{2}{c}{\rot{History}} & \multicolumn{2}{c}{\rot{Languages}} & \multicolumn{2}{c}{\rot{Law \& Order}} & \multicolumn{2}{c}{\rot{Media \& Culture}} \\ \midrule
Lexical & BM-25 & 0.63 &\textit{0.71} & \valbest{0.59} & \valbest{\textit{0.62}} &0.60 &\textit{0.70} &0.58 &\textit{0.69} &\valworst{0.62} &\valbest{\textit{0.62}} &0.58 &\textit{0.63} \\ \hline
\multirow{2}{*}{Sparse} & SPARTA & 0.63 & \valworst{\textit{0.74}} &0.61 & \textit{0.67} &\valbest{0.56} & \valbest{\textit{0.63}} &0.60 &\textit{0.67} &0.58 &\textit{0.67} &0.58 &\textit{0.66} \\
& SPLADE & 0.63 &\textit{0.68} &0.62 &\textit{0.67} &0.62 &\textit{0.67} &0.58 &\valworst{\textit{0.71}} &0.59 &\textit{0.66} &\valbest{0.57} &\textit{0.67} \\ \hline
\multirow{3}{*}{Dense} & ANCE & \valworst{0.69} &\textit{0.69} &0.63 & \valworst{\textit{0.68}} &\valworst{0.64} &\valworst{\textit{0.68}} &\valworst{0.62} &\textit{0.67} &\valworst{0.62} &\textit{0.64} &\valworst{0.61} &\textit{0.66} \\
& SBERT & 0.64 & \textit{0.67} & \valworst{0.64} &\textit{0.66} &0.60 &\textit{0.65} &0.60 &\textit{0.64} &0.58 &\textit{0.68} &0.60 &\textit{0.63} \\
& Use-QA & \valbest{0.58} & \textit{0.70} &0.60 &\valbest{\textit{0.62}} &0.57 &\textit{0.67} &\valbest{0.57} &\textit{0.65} &\valbest{0.56} &\textit{0.68} &\valbest{0.57} &\textit{0.63} \\ \hline
Late & ColBERT & 0.62 & \valbest{\textit{0.65}} &0.62 &\textit{0.65} &0.62 &\textit{0.66} &\valbest{0.57} &\textit{0.69} &{0.57} &\valworst{\textit{0.69}} &0.59 &\valworst{\textit{0.68}} \\ \midrule
Industry & Google & N/A & \textit{0.69} & N/A & \textit{0.66} & N/A & \textit{0.66} & N/A & \valbest{\textit{0.61}} & N/A & \textit{0.63} & N/A & \valbest{\textit{0.59}} \\
\bottomrule 
\multicolumn{4}{c}{\vspace{0.0005mm}}\\\toprule
Class & Model & \multicolumn{2}{c}{\rot{People}}
& \multicolumn{2}{c}{\rot{Politics \& Econ}} & \multicolumn{2}{c}{\rot{Psychiatry}}  
& \multicolumn{2}{c}{\rot{Science}} & \multicolumn{2}{c}{\rot{Sex \& Gender}} & \multicolumn{2}{c}{\rot{Technology}}  \\ \midrule
Lexical & BM-25 & 0.59 &\textit{0.67} &0.59 &\textit{0.67} &0.56 &\textit{0.65} &0.60 &\textit{0.68} &\valworst{0.59} &\textit{0.60} &0.58 &\valworst{\textit{0.76}} \\ \hline
\multirow{2}{*}{Sparse} & SPARTA & 0.59 &\textit{0.63} &0.58 &\textit{0.65} &0.55 &\valbest{\textit{0.52}} &\valbest{0.56} &\textit{0.60} &\valworst{0.59} &\textit{0.58} &0.59 &\valbest{\textit{0.65}} \\
& SPLADE & 0.60 &\textit{0.67} &0.60 &\textit{0.66} &0.53 &\textit{0.62} &0.62 &\valworst{\textit{0.69}} &\valworst{0.59} &\textit{0.59} &0.59 &\textit{0.72} \\ \hline
\multirow{3}{*}{Dense} & ANCE & \valworst{0.61} &\textit{0.67} &\valworst{0.61} &\valworst{\textit{0.71}} &0.65 &\textit{0.70} &\valworst{0.62} &\valworst{\textit{0.69}} &0.58 &\textit{0.61} &\valworst{0.61} &\textit{0.71} \\
& SBERT & 0.60 &\textit{0.66} &\valworst{0.61} &\textit{0.65} &\valworst{0.71} &\valworst{\textit{0.74}} &0.61 &\textit{0.66} &0.58 &\valworst{\textit{0.65}} &0.59 &\textit{0.74} \\
& Use-QA & \valbest{0.57} &\textit{0.64} &\valbest{0.56} &\textit{0.66} &\valbest{0.49} &\textit{0.68} &0.60 &\valbest{\textit{0.57}} &\valbest{0.53} &\textit{0.58} &\valbest{0.57} &\valbest{\textit{0.65}} \\ \hline
Late & ColBERT & 0.60 &\valworst{\textit{0.69}} &\valworst{0.61} &\textit{0.62} &0.56 &\textit{0.56} &0.61 &\textit{0.65} &0.58 &\valbest{\textit{0.54}} &\valbest{0.57} &\textit{0.70} \\ \midrule
Industry & Google & N/A & \valbest{\textit{0.59}} & N/A & \valbest{\textit{0.60}} & N/A & \textit{0.56} & N/A & \textit{0.60} & N/A & \textit{0.58} & N/A & \textit{0.67}\\
\bottomrule 
\end{tabular}
}
\caption{\textbf{Domain-Level \textsc{Wiki-Bias} results}  over \corpus{$_\text{Synthetic}$} (\textit{left columns}) and \corpus{$_\text{Natural}$} (\textit{italicized right columns}) can help identify entry points for critical bias-mitigation efforts. \textit{SBERT}, one of the most most overall biased models, specifically struggles with \textit{psychiatry}, \textit{entertainment}, and the \textit{environment}.}
\label{tab:domain_results}
\end{table*}

With \metric{} (\S\ref{sec:validation}) and evaluation corpora (\S\ref{sec:data}), we can audit the indexical bias of leading IR models. The labeled synthetic corpus shows us how \metric{} compares with prior bias metrics. However, synthetic results may not reflect end-model behavior on real document distributions. For more realistic results, we evaluate on the Natural corpus, where SEME behavioral experiments clearly validate \metric{} as predictive. The results in Tables~\ref{tab:aggregate_results} and \ref{tab:domain_results} use both standard relevance metrics and the bias metrics from \S\ref{sec:metrics}, and are averaged over 3 random seeds. Along with concurrent work \citep{zhao2024beyond}, \pair{} is among the first cross-domain audits of indexical biases in open-source IR systems.

\subsection{Models}  
Following BEIR \citep{thakur2021beir}, we use BM-25 as a strong lexical baseline, and evaluate six additional open-source neural systems, as well as one industrial system. For sparse models, we evaluate SPARTA \citep{zhao2020sparta} and SPLADE \citep{formal2021splade}. Next, we consider three dense models, ANCE \citep{xiong2020approximate}, SBERT \citep{reimers2019sentence}, and Use-QA \citep{yang2020neural}.
Our late-interaction model is ColBERT \citep{khattab2020colbert}. Finally, we evaluate Google's search engine. Our document embedding model is the best-performing \texttt{sentence-t5-xl}.

\subsection{Aggregate Bias Results} Here, \textbf{the most relevant models are not always the least biased.} Although SPLADE produces the most relevant results, it also introduces the most indexical bias in the natural evaluation setting (\metric{}$=0.62$). On the other hand, Use-QA is the least relevant model, yet it produces the least biased rankings in both the natural and synthetic evaluation (\metric$=0.64, 0.57$).

Most importantly, bias results validate our \metric{} metric, as \textbf{\metric{} highly correlates with the supervised metrics rND and rKL}, with Spearman correlations of 0.80 and 0.83 respectively ($p<0.05$). Thus \metric{} gives us similar conclusions about model bias without the need for human annotation. On the synthetic data, Use-QA and SPARTA are the least biased (\metric{}$\leq 0.58$; rKL $=0.60$), followed by ColBERT, SPLADE, and BM-25 (\metric{}$=0.60$). The most biased models are ANCE and SBERT (\metric{}$\geq 0.61$; rKL$\geq 0.62$). These results are stable; even if we compute \metric{} using a less accurate embedding model, the relative model order is roughly preserved ($\rho=0.72$; see Appendix~\ref{appdx:robustness_checks}).

\textbf{Model bias on the synthetic corpus also weakly predicts its bias on the natural corpus}, with a strong Spearman correlation of 0.64 between synthetic and natural \metric{} scores. ANCE remains the most biased model, while Use-QA remains the least biased open-source model. Google search is the least biased overall (\metric{} = 0.63). We conclude that synthetic evaluation may be used as a surrogate in this way to quickly evaluate IR systems across a wide range of domains. However, natural data remains the gold standard and should not be replaced by synthetic evaluations alone. Discrepancies between the synthetic and natural results may reflect differences between the distributions of their document polarizations. Synthetic data follows a highly-polarized bimodal distribution, while natural bias scores are both more neutral, and also normally distributed (see Appendix~\ref{appdx:differing_distributions}). The respective evaluations are mutually complementary.

Overall, \textbf{the orderings between models above are relatively stable}, even when we consider alternative embedding models, or when we apply debiasing methods to the embedding process to remove possible spurious correlations that arise from \corpus{$_\text{Synthetic}$} (see Appendix~\ref{appdx:robustness_checks} for more details on Experimental Replications, and Appendix~\ref{appdx:debiasing_embeddings} for methods to remove spurious artifacts).

\subsection{Aggregate Relevance Results} 
\textbf{Our relevance results confirm prior work} \citep{thakur2021beir}. Table~\ref{tab:aggregate_results} shows \textit{BM-25} is the strongest baseline for relevant retrieval on both \corpus{} corpora, and that \textit{BM-25} beats out models of greater complexity like \textit{ANCE} and \textit{SPARTA}. \textit{ColBERT} also achieves the top relevance scores on the \corpus{$_\text{Synthetic}$} corpus, which also aligns with \citet{thakur2021beir} and sanity-checks our results. 
The nDCG@10 relevance scores are all higher in the left side of the table, showing unsurprisingly that our synthetic corpus is an easier task than natural web retrieval.

\subsection{Domain-Level Results} One of \pair{'s} benefits is the ability to evaluate models domain-specific biases. Small aggregate differences in bias performance may not sway industry leaders and practitioners to adopt an entirely new IR system, but if an operational system demonstrates weakness in a particular domain, that can become a focal point for bias mitigation, like debiasing embeddings. Here, \textbf{\pair{} can serve as a precise instrument for diagnosing and addressing localized indexical biases}, as in this section.

Table~\ref{tab:domain_results} decomposes aggregate bias results into focal domains, revealing weaknesses in even the best models. The best open model, Use-QA, still falls short of Google Search in three key domains: \textsl{Psychiatry} (+0.12 more bias than Google), \textsl{Politics} (+0.06 \metric), and \textsl{Law} (+0.05 \metric). Since indexical bias in political search results can affect voting behavior \citep{epstein2015search}, practitioners may have strong incentives to mitigate such biases in open source systems.

In politics as in the aggregate, Google Search performs with the least bias on natural web data. However, given the power of any prominent search engine to influence countless users, it could be strategic for such search companies to invest greater attention towards bias mitigation at the weakest points. For Google, a weakness is the \textsl{Environment} (+0.04 more biased than the best Use-QA model).

\section{Applications and Extensions}
The \pair{} framework is general, and future work can explore its extensions outside of search in other domains where information follows a rank order, such as a chatbot's probability-ranked utterances, a politician's most common phrases, or even the ordered paragraphs of a news article \citep[i.e., \textit{framing};][]{ziems2021protect}. Indexical bias evaluation will become increasingly relevant to mitigate harms in more recent generative methods which rely on Retrieval Augmented Generation for grounding knowledge~\citep{lewis2020retrieval, shuster-etal-2022-language, jiang-etal-2023-active, khattab2022demonstrate}. 

A natural extension of the \metric{} could also handle issues with more than two sides, simultaneously incorporating multiple axes of semantic variation. Our formalization is readily prepared for such an extension. If we generate up to $m$ different viewpoint axes for each issue in \corpus{$_\text{Synthetic}$}, we could simply increase the dimensionality of the PCA projection in \S\ref{sec:metrics} such that polarization scores become polarization vectors $P_j \in \mathbb{R}^m$. This would allow a separate variance utility computation $u^{x}_v(i, r) = \frac{1}{i} \sum_{j=1}^{i} (P^{x}_j - \bar{P}^{x})^2$ and thus a separate score $\metric{}^x(r)$ for each viewpoint axis $x$. Depending on the application, one could aggregate across axes $x$, for example by taking the maximum bias score as in the multi-group Attention Bias Ratio of \citet{ghosh2021fair}. For more discussion of multi-group extensions of indexical bias metrics, see the Group Relevance Framework of \citet{sakai2023versatile}.

Additionally, there is still much to be learned about the formal and mathematical properties of the \metric{} metric, especially in the context of optimization and reranking. Prior works suggest that fluctuations in fairness metrics like \metric{} can lead to unstable training \citep{rekabsaz2021societal}. This may pose a challenge to the integration of \metric{} with current systems. One other bottleneck that may prevent the widespread adoption of \metric{} in reranking is the expensive computation for normalization. Our code includes a stochastic approximation, but the code could be further optimized.

\section{Conclusion}
From web search to personal assistants, IR systems have the potential to skew users' opinions on a wide range of topics, from media and entertainment to political issues and scientific insights. Before one can address the problematic outcomes of such manipulation, one should first expect to measure \textit{indexical bias}, the root of this psychological effect.
\pair{} is the first completely automatic method for evaluating indexical bias in IR systems without the need for manual human annotations. \pair{} is built on two bias evaluation corpora, which support our \metric{} metric. Since \metric{} requires no human supervision, it can serve as a scalable evaluation metric and, in future work, as an automatic reranking criterion. We demonstrate the psychological validity of \metric{} using a controlled experiment. After proving its validity, we use \metric{} to run an extensive audit over the biases in current IR technologies, both open and closed-source. Together, these contributions provide a basis for future efforts to measure and address indexical bias in IR. 

\section*{Limitations}
\label{sec:limitations_extensions}

\paragraph{Complementary Notions of Fairness.} All of the methodology we introduced in this work was focused on \textit{fairness of exposure} \citep{diaz2020evaluating}, balancing rankings to ensure \textit{equal visibility} \citep{pessach2022review} between groups or ideas. However, the appropriateness of fair exposure depends on the context \citep{singh2018fairness}. It may not always be socially desirable to balance certain viewpoints, as this may elevate or amplify hate, espouse misinformation, or jeopardize personal or collective well-being. For any practitioners interested in applying \pair{} or using the \corpus{} corpus, we strongly encourage a more careful selection of the topics and domains over which \metric{} balance is optimized.

Other related works have measured the complementary objective of \textit{fairness through neutrality} \citep{zerveas2022mitigating}, where systems are encouraged to preferentially retrieve more factual, neutral, and unbiased documents and to omit the most polarized documents entirely. Fairness through neutrality is critical and should not be ignored, especially for highly sensitive or ideological domains and settings where users may be most susceptible to confirmation bias \citep{del2017modeling} and its negative outcomes. 

Still, we note that \textit{fairness through neutrality} is not always attainable with a polarized corpus, nor is the notion of neutrality applicable to all queries \citep{krieg2023grep,zerveas2022mitigating}. For example, undecided voters may query a factual, encyclopediac corpus of Wikipedia articles on the biographies of candidates in an upcoming election. Voters opinions can shift by mere exposure to preferentially ranked biographies \citep{epstein2015search}. Here there is no truly neutral document, despite the academic, factual tone they carry.

On the other hand, \metric{} applies both to highly polarized and more neutral corpora. This is because \metric{} measures indexical bias as a relative quantity---how biased the ranked results are relative to the most imbalanced possible ranking of those same documents. In doing so, we disentangle system bias from any document biases in the corpus itself \citep{kulshrestha2017quantifying}.

\paragraph{Methodological Bias.} This work seeks to measure bias in IR systems. However, it is important to acknowledge potential biases in the \pair{} evaluation process itself. It is a non-trivial task to create an unbiased IR evaluation corpus. Just as traditional crowd-annotated datasets are prone to subjectivity biases in the document selection, relevance scoring, and other steps in the annotation pipeline, so also are synthetic methods vulnerable to such viewpoint biases, which may derive from the distribution of the LLM pretraining corpus, the prompt design, the seed topics used for prompting, or other related variables. The topics represented in \corpus{} were drawn automatically from English Wikipedia and extrapolated with language models. The authors of this paper did not hand-select any topics or documents, nor do we endorse any particular viewpoints in these resources. In the Appendix~\ref{appdx:scope}, we more thoroughly discuss biases in the seed topics, and we encourage future work to carefully consider and expand on this discussion.

\section*{Ethics}
This study has been approved by the Institutional Review Board (IRB) at the researchers' institution, and participant consent was obtained using the standard institutional consent form. For the annotation process, we included a warning in the instructions that the content might be offensive or upsetting. Annotators were also encouraged to stop the labeling process if they were overwhelmed, and regardless of how may tasks they completed, participants were paid a fair stipend of \$20 per hour for their time.

\section*{Acknowledgements}  We are thankful to the members of SALT Lab and the Stanford NLP Group for their helpful feedback on the draft. We especially appreciated suggestions from Tiziano Piccardi, Jing Huang, Faye Holt, Dora Zhao, Julia Kruk, and Michael Ryan. The work has been supported by a grant from Meta. Caleb Ziems is supported by the NSF Graduate Research Fellowship under Grant No. DGE-2039655. 

\bibliography{refs}

\appendix

\section{Appendix}
\label{sec:appendix}

\begin{figure}[h!]
    \centering \includegraphics[width=\linewidth]{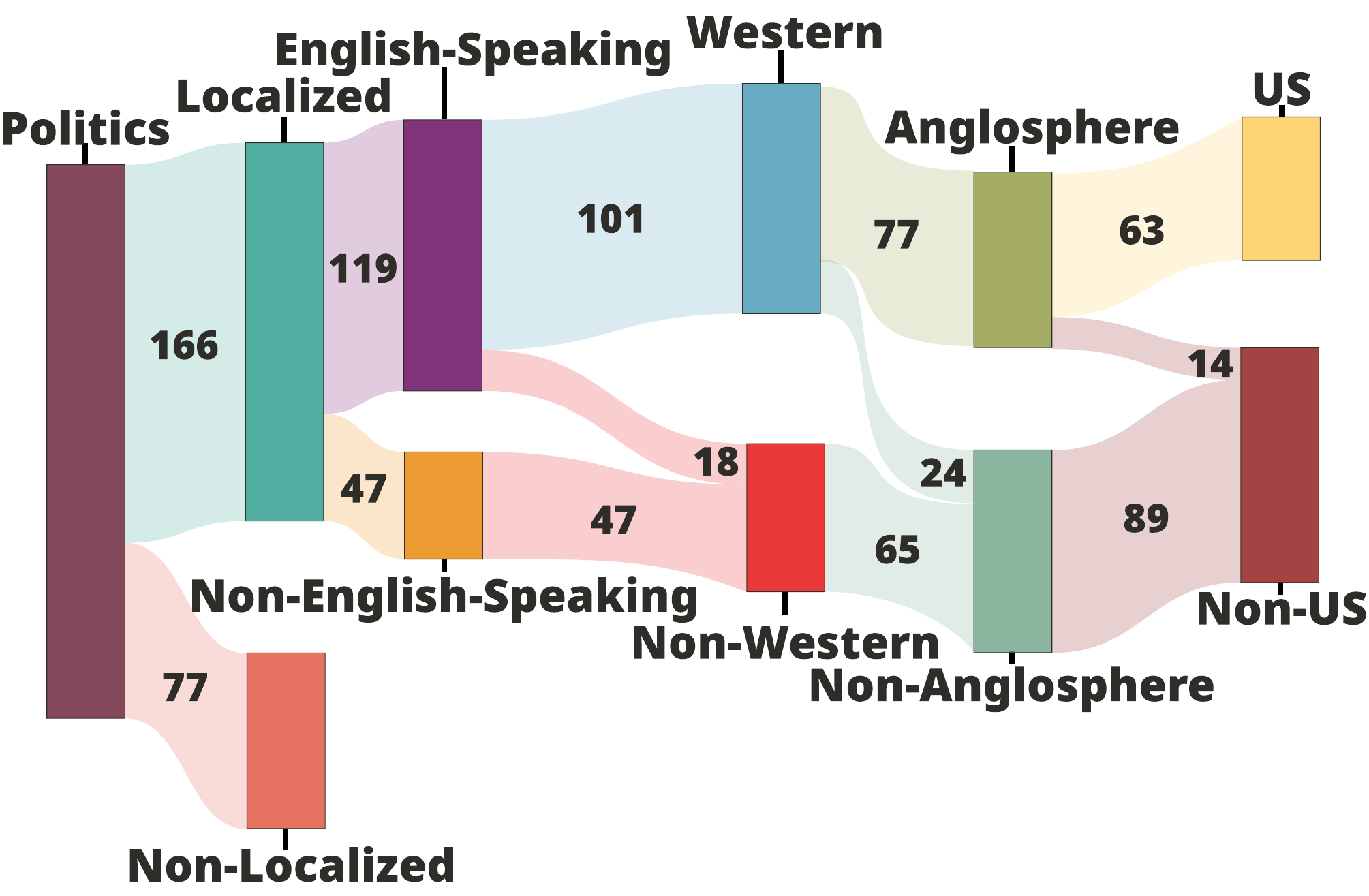}
    \caption{\textbf{Wikipedia Controversial Topic Distribution} can reflect biases in the Wikipedia editor pool. This explains why localized political topics are typically from English-speaking (71.69\%) countries, and why there is over-representation of American issues.}
    \label{fig:data_skew}
\end{figure}

\subsection{Potential Biases in \corpus{}}
\label{appdx:scope}

Controversial Wikipedia article titles are topically diverse, covering \numBiasDomains{} broad domains, including \textsl{politics, history, religion, science}. Given Wikipedia's global scope, we unsurprisingly find that each domain has wide coverage. Still, biases in the editor pool mean the distribution of topics in each domain is skewed. For example, in \textsl{politics}, local issues typically concern Anglophone countries (71.69\% of political topics are about countries in which English is either a official, majority, or secondary language). In Western politics, the United States is over-represented, appearing in 62\% of Western issues (see Figure~\ref{fig:data_skew}). Neither fact is surprising, since topics come from English Wikipedia, and a plurality English-language Wikipedians edit from the United States.\footnote{\tiny{\url{https://en.wikipedia.org/wiki/Wikipedia:Who_writes_Wikipedia}}} See \S\ref{sec:limitations_extensions} for a more in-depth discussion on the impacts of skewed data. 

\subsection{Measuring the Directionality of \metric{} Bias}  
\label{appdx:signed_ndcv}
By default, \metric{} metric measures only the magnitude of the bias and not its direction, as is the case with previous metrics \citep{gezici2021evaluation}.
Unlike prior metrics, our methodology allows for an unsupervised computation of a sign to indicate the \textit{direction} of the bias. If we consider the signum function of a real number
\begin{align*}
    \text{sgn}(x) = \begin{cases}
-1 & x < 0\\
0 & x = 0\\
1 & x > 0
\end{cases}
\end{align*}
we can extend this definition to a set $A \subset \mathbb{R}$ by

{\small  
  \setlength{\abovedisplayskip}{6pt}
  \setlength{\belowdisplayskip}{\abovedisplayskip}
  \setlength{\abovedisplayshortskip}{0pt}
  \setlength{\belowdisplayshortskip}{3pt}
\begin{align*}
    \text{sgn}(A) = 2\times\mathbb{1}\left[\left(\frac{|\{a_i \in A : a_i>0\}|}{|A|}\right)>0.5\right]+1
\end{align*}
}%
this gives us $\text{sgn}(A)=-1$ when $A$ contains more negative values than non-negative values, and $\text{sgn}(A)=1$ otherwise. If we modify the utility to encode the sign on the set of polarization scores $\{p_j\}$ as follows 
\begin{align*}
    u^{+}_{V}(i, r) = \text{sgn}(\{p_j\}_{j=1}^{i})u_V(i,r)
\end{align*}
Then we can set

{\small  
  \setlength{\abovedisplayskip}{6pt}
  \setlength{\belowdisplayskip}{\abovedisplayskip}
  \setlength{\abovedisplayshortskip}{0pt}
  \setlength{\belowdisplayshortskip}{3pt}
\begin{align*}
    \mu^{+}_{\metric{}}(r) = \text{sgn}\left(\metric{}(r, u^{+}_{V})\right)\left(\metric{}(r, u_V\right))
\end{align*}
}%
which assigns the existing \metric{} magnitude an appropriate polarity. Our signed $\mu^{+}_{\metric{}}(r)$ value will be useful for understanding how biased rankings can shift a reader's opinion \textit{towards} the perspective favored by the ranking (see \S\ref{subsec:validation_behavioral}).

\subsection{Expanded Accuracy of Polarization Embeddings}
\label{appdx:expanded_polarization_accuracies}

Here in Tables~\ref{tab:polarization_accuracy_results_above_90} and \ref{tab:polarization_accuracy_results_below_90}, we enumerate the unsupervised polarization label accuracy for each Transformer-based embedding model in the Sentence BERT library. Table~\ref{tab:polarization_accuracy_results_above_90} gives the 13 models with greater than 90\% accuracy, while Table~\ref{tab:polarization_accuracy_results_below_90} enumerates the remaining models in order of their accuracies. With a worst accuracy of 73\%, a best accuracy of 95\%, and a median accuracy of 87\%, these results strongly validate our approach, and show its robustness across document embedding implementation.

\begin{table}[ht!]
\small
\resizebox{\columnwidth}{!}{%
\def\arraystretch{1.15}
\begin{tabular}{ll}\toprule
\textbf{Embedding Model} & \textbf{Accuracy} \\ \midrule
\texttt{sentence-t5-xl} & 95.27\%\\ \hline
\texttt{sentence-t5-large} & 94.17\%\\ \hline
\texttt{nli-roberta-large} & 92.17\%\\ \hline
\texttt{roberta-large-nli-mean-tokens} & 92.17\%\\ \hline
\texttt{voyage-02} & 92.04\%\\ \hline
\texttt{paraphrase-distilroberta-base-v2} & 91.88\%\\ \hline
\texttt{paraphrase-mpnet-base-v2} & 90.86\%\\ \hline
\texttt{roberta-base-nli-stsb-mean-tokens} & 90.38\%\\ \hline
\texttt{roberta-large-nli-stsb-mean-tokens} & 90.29\%\\ \hline
\texttt{facebook-dpr-question\_encoder-single-nq-base} & 90.11\%\\ \hline
\texttt{roberta-base-nli-mean-tokens} & 90.05\%\\ \hline
\texttt{nli-roberta-base} & 90.05\%\\ \hline
\texttt{sentence-t5-base} & 90.03\%\\ \hline
\bottomrule 
\end{tabular}
}
\caption{\textbf{Sorted polarization score accuracies} for all embedding models in the Sentence BERT library \citep{reimers2019sentence} with accuracy greater than 90\%.}
\label{tab:polarization_accuracy_results_above_90}
\end{table}

\subsection{Differing Distributions: Natural And Synthetic}
\label{appdx:differing_distributions}
\begin{figure}
    \centering
    \begin{subfigure}{\columnwidth}
        \centering
        \includegraphics[width=0.7\columnwidth]{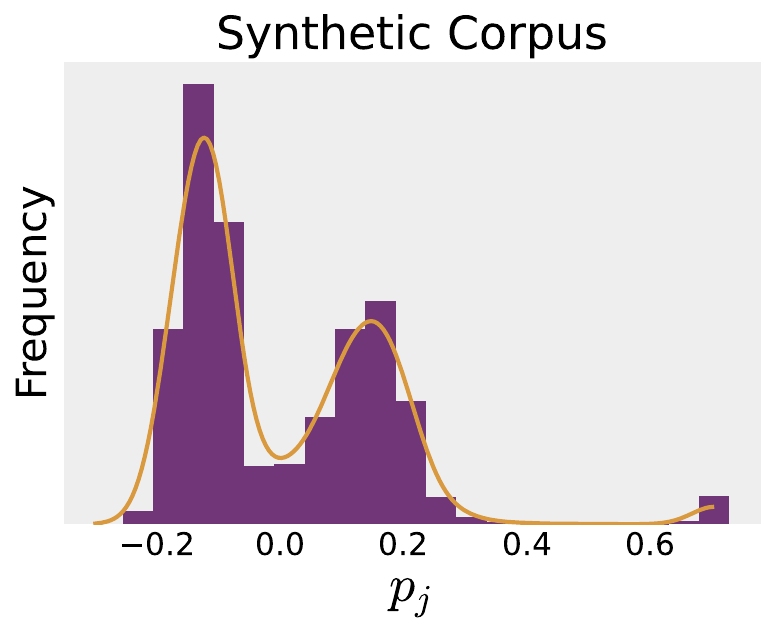}
    \end{subfigure}
    \begin{subfigure}{\columnwidth}
        \centering
        \includegraphics[width=0.7\columnwidth]{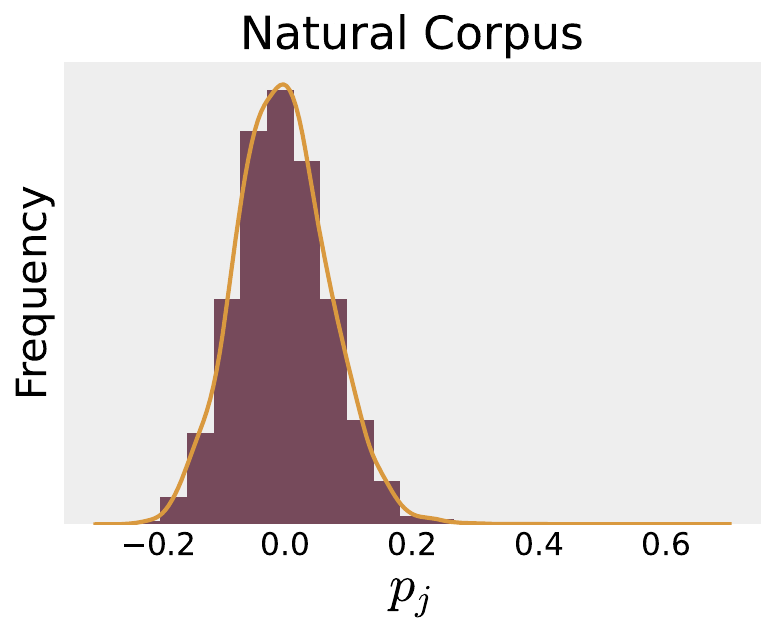}
    \end{subfigure}
    \caption{\textbf{Distributions of Polarization Scores} in the Synthetic (\textit{top}) and Natural (\textit{bottom}) corpus. Synthetic data is bimodal and polarized, while Natural data is normal and thus more neutral.}
    \label{fig:data_distributions}
\end{figure}

Unsurprisingly, Figure~\ref{fig:data_distributions} shows how natural Google search web articles are distributed differently than our synthetic corpus. Whereas natural bias scores (\textit{bottom}) are normally distributed around a neutral mean of zero, synthetic data (\textit{top}) follows a highly-polarized bimodal distribution, and includes some extreme outliers. We can conclude that any discrepancies between the synthetic and natural results in \S\ref{sec:bias_audits} are largely due to these differences. The respective evaluations are mutually complementary.

\begin{table}[ht!]
\small
\resizebox{\columnwidth}{!}{%
\def\arraystretch{1.15}
\begin{tabular}{ll}\toprule
\textbf{Embedding Model} & \textbf{Accuracy} \\ \midrule
\texttt{nli-bert-large-max-pooling} & 89.85\%\\ \hline
\texttt{bert-large-nli-max-tokens} & 89.85\%\\ \hline
\texttt{gtr-t5-large} & 89.70\%\\ \hline
\texttt{nli-bert-large} & 89.67\%\\ \hline
\texttt{bert-large-nli-mean-tokens} & 89.67\%\\ \hline
\texttt{gtr-t5-xl} & 89.58\%\\ \hline
\texttt{facebook-dpr-question\_encoder-multiset-base} & 89.39\%\\ \hline
\texttt{bert-large-nli-stsb-mean-tokens} & 89.28\%\\ \hline
\texttt{nli-bert-large-cls-pooling} & 89.10\%\\ \hline
\texttt{bert-large-nli-cls-token} & 89.10\%\\ \hline
\texttt{average\_word\_embeddings\_komninos} & 88.30\%\\ \hline
\texttt{average\_word\_embeddings\_glove.6B.300d} & 88.29\%\\ \hline
\texttt{LaBSE} & 88.27\%\\ \hline
\texttt{msmarco-roberta-base-ance-firstp} & 88.24\%\\ \hline
\texttt{paraphrase-MiniLM-L12-v2} & 87.78\%\\ \hline
\texttt{average\_word\_embeddings\_levy\_dependency} & 87.78\%\\ \hline
\texttt{nli-bert-base} & 87.77\%\\ \hline
\texttt{bert-base-nli-mean-tokens} & 87.77\%\\ \hline
\texttt{distilbert-base-nli-stsb-mean-tokens} & 87.70\%\\ \hline
\texttt{bert-base-nli-stsb-mean-tokens} & 87.59\%\\ \hline
\texttt{nli-bert-base-max-pooling} & 87.41\%\\ \hline
\texttt{bert-base-nli-max-tokens} & 87.41\%\\ \hline
\texttt{nli-roberta-base-v2} & 87.40\%\\ \hline
\texttt{average\_word\_embeddings\_glove.840B.300d} & 87.33\%\\ \hline
\texttt{gtr-t5-base} & 87.25\%\\ \hline
\texttt{nli-bert-base-cls-pooling} & 87.17\%\\ \hline
\texttt{bert-base-nli-cls-token} & 87.17\%\\ \hline
\texttt{all-mpnet-base-v1} & 87.03\%\\ \hline
\texttt{nli-distilbert-base-max-pooling} & 86.98\%\\ \hline
\texttt{distilbert-base-nli-max-tokens} & 86.98\%\\ \hline
\texttt{msmarco-distilbert-base-tas-b} & 86.94\%\\ \hline
\texttt{nli-mpnet-base-v2} & 86.94\%\\ \hline
\texttt{nli-distilbert-base} & 86.78\%\\ \hline
\texttt{distilbert-base-nli-mean-tokens} & 86.78\%\\ \hline
\texttt{msmarco-roberta-base-v3} & 86.53\%\\ \hline
\texttt{multi-qa-mpnet-base-dot-v1} & 86.40\%\\ \hline
\texttt{all-distilroberta-v1} & 86.24\%\\ \hline
\texttt{paraphrase-distilroberta-base-v1} & 85.88\%\\ \hline
\texttt{distilroberta-base-paraphrase-v1} & 85.88\%\\ \hline
\texttt{multi-qa-mpnet-base-cos-v1} & 85.53\%\\ \hline
\texttt{msmarco-bert-base-dot-v5} & 85.32\%\\ \hline
\texttt{nli-distilroberta-base-v2} & 85.01\%\\ \hline
\texttt{msmarco-MiniLM-L-12-v3} & 84.93\%\\ \hline
\texttt{paraphrase-xlm-r-multilingual-v1} & 84.87\%\\ \hline
\texttt{msmarco-distilbert-dot-v5} & 84.65\%\\ \hline
\texttt{nq-distilbert-base-v1} & 84.45\%\\ \hline
\texttt{msmarco-MiniLM-L12-cos-v5} & 84.33\%\\ \hline
\texttt{msmarco-roberta-base-v2} & 84.32\%\\ \hline
\texttt{msmarco-distilbert-base-v3} & 84.31\%\\ \hline
\texttt{multi-qa-distilbert-cos-v1} & 84.12\%\\ \hline
\texttt{msmarco-distilbert-base-dot-prod-v3} & 84.08\%\\ \hline
\texttt{paraphrase-MiniLM-L6-v2} & 83.90\%\\ \hline
\texttt{msmarco-distilroberta-base-v2} & 83.87\%\\ \hline
\texttt{distilroberta-base-msmarco-v2} & 83.87\%\\ \hline
\texttt{all-MiniLM-L6-v2} & 83.84\%\\ \hline
\texttt{msmarco-distilbert-base-v2} & 83.76\%\\ \hline
\texttt{msmarco-distilbert-multilingual-en-de-v2-tmp-lng-aligned} & 83.76\%\\ \hline
\texttt{paraphrase-multilingual-mpnet-base-v2} & 83.41\%\\ \hline
\texttt{multi-qa-distilbert-dot-v1} & 83.33\%\\ \hline
\texttt{msmarco-distilbert-base-v4} & 83.16\%\\ \hline
\texttt{msmarco-bert-co-condensor} & 83.13\%\\ \hline
\texttt{msmarco-distilbert-cos-v5} & 83.12\%\\ \hline
\texttt{msmarco-MiniLM-L-6-v3} & 83.08\%\\ \hline
\texttt{msmarco-MiniLM-L6-cos-v5} & 82.99\%\\ \hline
\texttt{multi-qa-MiniLM-L6-cos-v1} & 82.90\%\\ \hline
\texttt{all-mpnet-base-v2} & 82.82\%\\ \hline
\texttt{all-MiniLM-L12-v1} & 82.81\%\\ \hline
\texttt{facebook-dpr-ctx\_encoder-single-nq-base} & 82.71\%\\ \hline
\texttt{paraphrase-albert-base-v2} & 82.71\%\\ \hline
\texttt{paraphrase-TinyBERT-L6-v2} & 82.70\%\\ \hline
\texttt{distilroberta-base-msmarco-v1} & 82.61\%\\ \hline
\texttt{bert-base-wikipedia-sections-mean-tokens} & 81.70\%\\ \hline
\texttt{multi-qa-MiniLM-L6-dot-v1} & 81.46\%\\ \hline
\texttt{distiluse-base-multilingual-cased-v2} & 81.43\%\\ \hline
\texttt{distiluse-base-multilingual-cased} & 81.43\%\\ \hline
\texttt{all-roberta-large-v1} & 81.19\%\\ \hline
\texttt{paraphrase-multilingual-MiniLM-L12-v2} & 81.09\%\\ \hline
\texttt{paraphrase-MiniLM-L3-v2} & 80.78\%\\ \hline
\texttt{all-MiniLM-L6-v1} & 80.77\%\\ \hline
\texttt{paraphrase-albert-small-v2} & 80.66\%\\ \hline
\texttt{msmarco-distilbert-multilingual-en-de-v2-tmp-trained-scratch} & 80.47\%\\ \hline
\texttt{quora-distilbert-base} & 80.40\%\\ \hline
\texttt{distilbert-base-nli-stsb-quora-ranking} & 80.40\%\\ \hline
\texttt{distilbert-multilingual-nli-stsb-quora-ranking} & 80.33\%\\ \hline
\texttt{quora-distilbert-multilingual} & 80.33\%\\ \hline
\texttt{allenai-specter} & 80.05\%\\ \hline
\texttt{all-MiniLM-L12-v2} & 79.49\%\\ \hline
\texttt{distiluse-base-multilingual-cased-v1} & 78.89\%\\ \hline
\texttt{facebook-dpr-ctx\_encoder-multiset-base} & 78.12\%\\ \hline
\texttt{clip-ViT-B-32-multilingual-v1} & 73.19\%\\ \hline
\bottomrule 
\end{tabular}
}
\caption{\textbf{Sorted polarization score accuracies} for all embedding models in the Sentence BERT library \citep{reimers2019sentence} with accuracy lower than 90\%.}
\label{tab:polarization_accuracy_results_below_90}
\end{table}

\subsection{Removing Spurious Correlations from Polarization Embeddings}
\label{appdx:debiasing_embeddings}

\metric{} depends on \corpus{$_\text{Synthetic}$} for the $|r|$ synthetic documents on which polarization scores $p_j$ are computed via PCA. The synthetic data can introduce spurious correlations into the \metric{} metric. For example, with the following query
\begin{quote}
    \textit{How can the legal system effectively enforce copyright laws in BitTorrent-enabled piracy?}
\end{quote}
we have two perspectives: (1) \textit{from the perspective of the Entertainment Industry}, and (2) \textit{from the perspective of BitTorrent users}. Naturally, any synthetic documents generated for Perspective 1 will contain more legal language. This means that even neutral documents could be marked by our embedding method as supporting Perspective 1 if they contain legal language (e.g., a purportedly neutral Wikipedia article). Here, we will describe how we worked to understand and remove such spurious correlations in \metric{}. 

Following prior works on the removal of bias in word embeddings \citep{bolukbasi2016man,lauscher2020general}, we opt to identify in the document embedding space some subspace in which the spurious correlation lies; then we can effectively project away this subspace. Let $b$ be the principle bias axis in our embedding space, and $x_j$ be the raw document embedding for document $j$. The debiased document embedding will then be
\begin{align*}
    \Tilde{x}_j = x_j - \left<x_j, b\right>b
\end{align*}
As before, we can fit PCA on $\Tilde{X} = [\Tilde{x}_1; ...; \Tilde{x}_8]$ to compute de-biased polarization scores $\Tilde{p}_j$. 

To identify $b$, we used the following process, extracting the axis of spurious correlation automatically from a set of ``distractor documents.'' First we generate these distractor documents to represent spurious correlations in the data. These distractors are the outputs of a pipeline similar to that of \S\ref{subsec:data_idea_balance} for creating \corpus{$_\text{Synthetic}$}. For a given document $d^i_t$, which answers some query $q_t$ (e.g., ``\textit{Did Edison steal patents from Tesla?}'') about a high-level
topic $t$ (e.g., \textit{Nikola Tesla}) by endorsing one particular perspective $P^i_t$ (e.g., ``\textit{Edison stole patents from Tesla.}''), we will generate a distractor document $\Tilde{d}^i_t$ that use similar style and vocabulary as $d^i_t$ without ever answering $q_t$. The document $\Tilde{d}^i_t$ will neither endorse nor deny $P^i_t$. To this end, we retrieve a distractor question $\Tilde{q}^i_t$, which is a query semantically distinct from ${q}^i_t$, but it concerns the same high-level topic (e.g., the query, ``\textit{How did Tesla's personal eccentricities influence his work?}''). With the distractor question $\Tilde{q}^i_t$, we generate $\Tilde{d}^i_t$ by prompting \texttt{gpt-3.5-turbo}, \textbf{``\textit{<$d^i_t$> Using as many words and phrases from the paragraph above as possible, try to answer: <$\Tilde{q}^i_t$>}.''} To ensure that $\Tilde{d}^i_t$ is neutral with respect to $q_t$, we follow up with, \textbf{``Rewrite the above paragraph but remove any sentences that have to do with the idea: <$P^i_t$>.''} 

Now for each original document $d^i_t$ we have a distractor $\Tilde{d}^i_t$. Now we want to identify a spurious decision boundary between each perspective, so we partition the distractors according to the perspective of the document they stylistically emulate: 
\begin{align*}
    D_1 = \{\Tilde{d}^i_t : P^i_t=1\} & &
    D_2 = \{\Tilde{d}^i_t : P^i_t=2\}
\end{align*}
If each set of documents has its own average document embedding $\mu_{D_1}, \mu_{D_2}$, we can set the principle axis of spurious correlation to be
\begin{align*}
    b = \mu_{D_1} - \mu_{D_2}
\end{align*}
This difference of means is a proven reliable method for identifying the semantic direction between binary concepts \citep{marks2023geometry}.

\subsection{Experimental Replications}  
\label{appdx:robustness_checks}
In this section, we replicate our studies with different parameters to demonstrate the robustness of our experimental results in both the SEME Behavioral Study in \S\ref{subsec:validation_behavioral} and the Bias Audit in \S\ref{sec:bias_audits}. 

\paragraph{Bias Audit.} Table~\ref{tab:replication_results} gives the experimental replications for the Bias Audit, using different embedding methods. The \textsl{MiniLM} columns indicate the use of \texttt{all-MiniLM-L6-v2}, a weaker embedding model, as its accuracy in Table~\ref{tab:polarization_accuracy_results_below_90} was only 83\%, compared with the stronger 95\% performance of \texttt{sentence-t5-xl}. The \textsl{MiniLM} synthetic results have a correlation of $\rho=0.72$ with the T5-XL results. Separately, in the \textsl{T5-XL-Debiased} column, we replicate our findings using the debiasing methods from \S\ref{appdx:debiasing_embeddings} and find a strong correlation of $\rho=0.89$ with the synthetic \textsl{T5-XL} results, and $\rho=0.94$ with the natural \textsl{T5-XL} results. In each case, debiasing preserves the relative model ordering at the bottom of Table~\ref{tab:replication_results}. For example, on \corpus{$_\text{Synthetic}$}, we have Use-QA $\succ$ \textsl{SPARTA} $\succeq$ \textsl{BM-25} $\succ$ \textsl{ColBERT} $\succeq$ \textsl{SPLADE} $\succ$ \textsl{SBERT} $\succ$ \textsl{ANCE}. We can conclude that our \pair{} methodology largely induces a stable relative model ordering from the \metric{} metric, even when we consider alternative embedding models or apply debiasing methods to the embedding process to remove any potential spurious correlations.

\paragraph{SEME.} Finally, we run experimental replications for the SEME in which we try both different embedding systems and different \metric{} step sizes. In Table~\ref{tab:polarization_accuracy_results_below_90}, we determined that \texttt{clip-ViT-B-32-multilingual-v1} gives the lowest polarization score accuracy of 73\%. Now in Table~\ref{tab:seme_regression_results_robustness}, we find that this polarization accuracy is not high enough to produce a significant effect in the SEME experiment. However, we can replicate our significant findings for a mid-performance model, \texttt{all-MiniLM-L6-v2}, which has 84\% polarization accuracy, and the top-performance model, \texttt{sentence-t5-xl} which has 95\% polarization accuracy. We can also replicate these experiments when we increase the \metric{} step size from 1 to 2 as in prior work \citep{yang2017measuring}. To increase the step size, one can effectively substitute the DCG in Equation~\ref{eq:dcg} with
\begin{equation*}
\text{DCG}(r, u) = \sum_{i=2, 4, ...}^{|r|}\frac{u(i, r)}{\log_2{i}}    
\end{equation*}
where $i$ increments in steps of 2. In all such replications, we observe significant ($p<0.05$) positive $\beta_2$ coefficients, leading us to conclude that \metric{} helps predict the SEME in cases of article click-through, and thus validating our method.  

\begin{table}[t!]
\small
\resizebox{\textwidth}{!}{%
\def\arraystretch{1.15}
\begin{tabular}{ll|ccccccc}\toprule
Class & Model & \multicolumn{2}{c}{MiniLM} & \multicolumn{2}{c}{T5-XL} & \multicolumn{2}{l}{T5-XL Debiased} \\ \midrule
Lexical & BM-25& 0.67 & \textit{0.64} & 0.60 & \textit{0.66} & 0.60 & \textit{0.67}\\ \hline
\multirow{2}{*}{Sparse} & SPARTA& 0.65 & \textit{0.65} & 0.58 & \textit{0.65} & 0.60 & \textit{0.65} \\
& SPLADE& 0.68 & \textit{0.68} & 0.60 & \textit{0.67} & 0.61 & \textit{0.67}\\ \hline
\multirow{3}{*}{Dense} & ANCE& 0.70 & \textit{0.67} & 0.62 & \textit{0.67} & 0.63 & \textit{0.67}\\
& SBERT&  0.70 & \textit{0.69} & 0.61 & \textit{0.65} & 0.61 & \textit{0.65} \\
& Use-QA& 0.68 & \textit{0.64} & 0.57 & \textit{0.64} & 0.58 & \textit{0.63}\\ \hline
Late & ColBERT& 0.68 & \textit{0.67} & 0.60 & \textit{0.65} & 0.61 & \textit{0.64} \\ \midrule \midrule
\multicolumn{9}{c}{Relative Model Orderings (Synthetic)} \\ \midrule
MiniLM: & \multicolumn{8}{l}{SPARTA $\succ$ Use-QA $\succeq$ BM-25 $\succeq$ ColBERT $\succeq$ SPLADE $\succ$ SBERT $\succeq$ ANCE}\\
T5-XL: & \multicolumn{8}{l}{Use-QA $\succ$ SPARTA $\succeq$ BM-25 $\succeq$ ColBERT $\succeq$ SPLADE $\succ$ SBERT $\succ$ ANCE}\\
T5-XL Debiased: & \multicolumn{8}{l}{Use-QA $\succ$ SPARTA $\succeq$ BM-25 $\succ$ ColBERT $\succeq$ SPLADE $\succeq$ SBERT $\succ$ ANCE}\\ \midrule
\multicolumn{9}{c}{Relative Model Orderings (\textit{Natural})} \\ \midrule
MiniLM: & \multicolumn{8}{l}{Use-QA $\succeq$ BM-25 $\succ$ SPARTA $\succ$ ColBERT $\succeq$ ANCE $\succ$ SPLADE $\succ$ SBERT}\\
T5-XL: & \multicolumn{8}{l}{Use-QA $\succ$ ColBERT $\succeq$ SPARTA $\succeq$ SBERT $\succ$ BM-25 $\succ$ SPLADE $\succ$ ANCE}\\
T5-XL Debiased: & \multicolumn{8}{l}{Use-QA $\succ$ ColBERT $\succ$ SPARTA $\succeq$ SBERT $\succ$ BM-25 $\succeq$ SPLADE $\succeq$ ANCE}\\
\bottomrule 
\end{tabular}
}
\caption{\textbf{Experimental Replications of the Bias Audit} over \corpus{$_\text{Synthetic}$} (\textit{left columns}) and \corpus{$_\text{Natural}$} (\textit{italicized right columns}). Here we report \metric{} using different embedding methods: (1) using a weaker embedding model (\textsl{MiniLM}), and (2) using the debiasing method from \S\ref{appdx:debiasing_embeddings} (\textsl{T5-XL-Debiased}). Scores have high mutual correlation, and relative model orderings (\textit{bottom}) are stable across these replications.}
\label{tab:replication_results}
\end{table}

\begin{table*}[t!]
\small
\resizebox{\columnwidth}{!}{%
\def\arraystretch{1.15}
\begin{tabular}{lrcccccc}\toprule
\textbf{Corpus} & \textbf{Behavior} 
& Step & \metric{} Emb Model & $N$ & $\beta_2$ & $P(\beta_2=0)$ & \textbf{$R^2$} \\ \midrule
Natural & Clicked & 1 & clip-ViT-B-32-multilingual-v1 & 99 & -0.294 & 0.176 & 0.475\\
\rowcolor{valbest} Natural & Clicked & 1 & all-MiniLM-L6-v2 & 99 & 0.394 & 0.036 & 0.489  \\
\rowcolor{valbest} Natural & Clicked & 1 & sentence-t5-xl & 99 & 0.527 & 0.022 & 0.493\\
Natural & Clicked & 2 & clip-ViT-B-32-multilingual-v1 & 99 & -0.271 & 0.192 & 0.474\\
\rowcolor{valbest} Natural & Clicked & 2 & all-MiniLM-L6-v2 & 99 & 0.392 & 0.036 & 0.489\\
\rowcolor{valbest} Natural & Clicked & 2 & sentence-t5-xl & 99 & 0.492 & 0.025 & 0.492\\
Combined & Clicked & 1 & clip-ViT-B-32-multilingual-v1 & 118 & -0.173 & 0.373 & 0.503 \\
\rowcolor{valbest} Combined & Clicked & 1 & all-MiniLM-L6-v2 & 118 & 0.366 & 0.032 & 0.519 \\
\rowcolor{valbest} Combined & Clicked & 1 & sentence-t5-xl & 118 & 0.426 & 0.046 & 0.517\\
Combined & Clicked & 2 & clip-ViT-B-32-multilingual-v1 & 118 & -0.164 & 0.377 & 0.503\\
\rowcolor{valbest} Combined & Clicked & 2 & all-MiniLM-L6-v2 & 118 & 0.365 & 0.032 & 0.519\\
\rowcolor{valbest} Combined & Clicked & 2 & sentence-t5-xl & 118 & 0.393 & 0.051 & 0.516\\
\bottomrule 
\end{tabular}
}
\caption{\textbf{Experimental Replications of the SEME Behavioral Study} with different embedding systems. The weakest embedding model, \texttt{clip-ViT-B-32-multilingual-v1} fails to produce a significant effect, due to its low polarization score accuracy of 73\% as determined in Table~\ref{tab:polarization_accuracy_results_below_90}. However, we can replicate our significant findings for a mid-performance model, \texttt{all-MiniLM-L6-v2}, which has 84\% polarization accuracy, and a top-performance model, \texttt{sentence-t5-xl} which has 95\% polarization accuracy. We can also replicate these experiments when we increase the \metric{} step size from 1 to 2 as in prior work \citep{yang2017measuring}. In all such replications, we observe significant ($p<0.05$) positive $\beta_2$ coefficients, leading us to conclude that \metric{} helps predict the SEME in cases of article click-through, and thus validating our method.}
\label{tab:seme_regression_results_robustness}
\end{table*}

\subsection{Parameters and Computing Budget}
All experiments were performed on a Ubuntu Linux machine with 6 Nvidia GeForce RTX 2080 Ti GPUs. Each model evaluation took around 4 hours. All embedding models were run with the Sentence BERT \citep{reimers2019sentence} package on default parameters. All IR models were run with the BEIR \citep{thakur2021beir} package on default parameters.

\end{document}